\documentclass[fleqn,usenatbib]{mnras}

\usepackage{newtxtext,newtxmath}

\usepackage[T1]{fontenc}
\usepackage{ulem}
\DeclareRobustCommand{\VAN}[3]{#2}
\let\VANthebibliography\thebibliography
\def\thebibliography{\DeclareRobustCommand{\VAN}[3]{##3}\VANthebibliography}

\usepackage{graphicx}	
\usepackage{amsmath}	

\title{Compact Groups in GDM Cosmological Simulations}
\author[Jessica N. López-Sánchez et al.]{
Jessica N. López-Sánchez,$^{1,2}$\thanks{E-mail: jessica.lopezsan@alumno.buap.mx}
Erick Munive-Villa,$^{1,2}$\thanks{E-mail: erick.munive@alumno.buap.mx}
Ana Avilez-López,$^{1,2}$\thanks{E-mail: ana.avilezlopez@correo.buap.mx}
Oscar M. Martínez-Bravo,$^{1,2}$\thanks{E-mail: oscar.martinezb@correo.buap.mx}
\\
$^{1}$Facultad de Ciencias F\'isico-Matem\'aticas, Ciudad Universitaria, Benem\'erita Universidad Aut\'onoma de Puebla,\\ Av. San Claudio SN, Col. San Manuel, Puebla, M\'exico\\
$^{2}$Centro Intenacional de F\'isica Fundamental\\
Ciudad Universitaria, Benem\'erita Universidad Aut\'onoma de Puebla,\\ Av. San Claudio SN, Col. San Manuel, Puebla, M\'exico\\
}

\date{Accepted XXX. Received YYY; in original form ZZZ}

\pubyear{2022}

\begin{document}
\label{firstpage}
\pagerange{\pageref{firstpage}--\pageref{lastpage}}
\maketitle

\begin{abstract}
In this work, we study some properties of the Hickson Compact Groups (HCGs) using N-body simulations for the Generalized Dark Matter (GDM) model, described by three free functions, the sound speed, the viscosity and the equation of state. We consider three GDM models associated with different values of the free functions to neglect collisional effects. We constructed the initial seeds of our simulations according to the matter power spectrum of GDM linear perturbations, which hold a cut-off at small scales, and explored their effects on the non-linear structure formation at small and intermediate scales. We generated mock catalogues of galaxies for different models and classify HCGs by implementing an algorithm that adapts the original selection method for mock catalogues. Once the HCGs samples are classified, we analyzed their properties and compared them between models. We found that a larger amount of HCGs are counted in GDM simulations in comparison to CDM counts. This difference suggests that HCGs can proliferate within GDM despite the suppressed substructure, which indicates a possible modification in the HCG formation process within models where DM is not perfectly like CDM. Additionally, we identified different mechanisms of clustering, for models with a large amount of galaxy-halos self-agglomerate because of their abundance while models with fewer galaxy-halos need massive halos acting as a dominant potential well. Finally, by comparing distributions of different observables of simulated HCGs against observations, we found a good agreement in the intrinsic properties. However, a discrepancy in the velocity dispersion remains unsolved.
\end{abstract}



\maketitle

\section{Introduction}
\label{sec:1}
\subsection{Dark Matter as a general fluid}
\label{sec:1a}
One of the most notable successes of the Standard Model of Cosmology $\Lambda$CDM has been the right prediction of the accelerated expansion of the universe, the description of the structure formation at large scales at late times in the expansion history and the stunning precise prediction of the temperature and polarisation anisotropies spectra of cosmic background radiation decoupled from the primordial plasma at the early times \citep{spergel}.

Observations of rotation curves of galaxies \citep{Vera_Rubin} as well as the dynamics of clusters \citep{Yang} suggest that dark matter has been presented at different scales and ages of the universe. Moreover, observations of the distribution of matter in the bullet cluster \citep{Barrena}, measurements of the anisotropies of the CMB and the study of the structure at large and intermediate scales \citep{Blumenthal} suggest this component mainly interacts with other species and itself gravitationally. For those reasons, within the standard $\Lambda$CDM paradigm, dark matter is modelled as a non-collisional perfect fluid.

However, some aspects give rise to some tension within the $\Lambda$CDM model, such as the cosmological constant fine-tuning problem \citep{Carroll1992}, the statistical anomalies due to the large angle fluctuations in the CMB \citep{Sugiyama}, the actual values of the Hubble parameter $H_0$ and $\sigma_8$ \citep{Verde:2019, Planck2018}. Specifically, regarding CDM piece of the standard model, there are still open questions when describing structures at small scales \citep{ostriker,kolchin,moore-1}. For instance, the cusp-core problem corresponds to a discrepancy between the flat density profiles of dwarf and low surface brightness (LSB) galaxies inferred from observations \citep{van-den,Shapiro} and the cuspy profile arising in N-body simulations within CDM \citep{moore-1,Flores,McGauch}. Also, there exists the missing satellite problem, coining the discrepancy between the number of predicted sub-halos in N-body simulations and those observed around the Milky Way in the Local Group and it is related to the Too Big To Fail problem \citep{kolchin}, due to in $\Lambda$CDM prediction, satellites are too massive and dense compared to the observations \citep{Moore-1999,Klypin}.

There have been different proposals to solve one or more of the mentioned problems such as the inclusion of baryonic physics or the modifications to the Standard Model, for instance in the dark matter sector. Currently, there are some candidates for which the assumption of being a pressureless perfect fluid is softly relaxed to consider a more general approach. There are models which arise from a purely phenomenological description of a fluid allowing non-vanishing pressure and shear encompassed within a general framework dubbed as the GDM model (GDM) \citep{kopp,Hu}.

GDM is a purely phenomenological approach constructed to constrain the properties of dark matter within the linear regime and was first introduced by Wayne Hu in \citet{Hu}. GDM is intended to encompass a large class of models in which dark matter is described as a general fluid using three free functions associated with their general properties. Specifically, it contains one time-dependent free function $w(a)=\bar{P}_g/\bar{\rho}_g$ (the equation of state) and two free functions $c_{\text{s}}^2(k,a)$ (the sound speed) and $c_{\text{vis}}^2(k,a)$ (the viscosity), which depend on scale $k$ as well as the scale factor $a$, but are solution-independent \citep{kopp}. An interesting feature about GDM is its capability of reproducing some other well-known models such as Hot Dark Matter (HDM) \citep{Primack, Bond-Centrella} or Scalar Field Dark Matter (SFDM) (also dubbed as Fuzzy dark matter (FDM)) \citep{Hui,Matos_2000,Hu2000} and of course CDM \citep{Hu}.

In the limit where $(w, c_{\text{s}}^2, c_{\text{vis}}^2)\rightarrow (1/3,1/3,1/3)$ HDM is reproduced, whereas $(w, c_{\text{s}}^2, c_{\text{vis}}^2)\rightarrow (w,1,0)$ corresponds to SFDM and $(w, c_{\text{s}}^2, c_{\text{vis}}^2)\rightarrow (0,0,0)$ to CDM \citep{kopp}. Hence, in order to study a scenario close to CDM it is necessary to consider $|w|$, $c_{\text{s}}^2$, $c_{\text{vis}}^2\ll 1$. The values for the free parameters determine the fluid properties and thereof the dynamics of the linear perturbations which bring up effects on the Cosmic Microwave Background (CMB) anistropies and the Matter Power Spectrum describing structures at large scales in the universe.

\subsection{Galaxy populations in different DM models}
\label{sec:1b}
Since structure formation at small and intermediate scales as well as galaxy and cluster formation and dynamics strongly depend on the properties of dark matter, phenomena involved in these highly non-linear processes, provide a large laboratory to study the properties of dark matter. In the literature, a large amount of information about populations of galaxies through time and space, in the CDM paradigm, has arisen from N-body simulations\citep{illustris,Busha,Guo,Delucia-sim,Bower}. 
Information about the physical properties of the galaxies and the halos where they reside makes it possible to examine the galaxy population and the way they cluster, for instance, when forming rich groups, loose groups or compact groups of galaxies (CGs).

In this work, we particularly focus on studying the CG of galaxies, which are agglomerations of a few galaxies within a small region in the sky considered to have a high density and velocity dispersion \citep{1995Pildis,1995PildisA,1996Pildis,1995Saracco}. These factors combined give rise to a very short time scale for the dynamical evolution of such groups whose consequence is believed to be the merger of all the bound galaxies into a single and more stable system \citep{hickson-cg}. Currently, there is still no explanation about the current number of CGs in existence, although several theories have been proposed \citep{tovmassian-2006, mamon_2000, Ramella, Vennik1993}. 

Furthermore, several studies of mock catalogues derived from the Millennium Simulation \citep{NGENICa} have been used to identify and analyse CGs. In \citet{McConnachie} a mock catalogue was used to investigate the spatial properties of CGs, they found that around $30$ per cent of galaxy associations are physically dense systems and more than half of the groups identified consist of a single dark matter halo with all the galaxies members embedded. On the other hand, in \citet{Gimenez-Mamon} CGs were identified using a 2D selection criteria and then they determined the fraction of physically dense groups using the maximum physical separation between galaxies, concluding more than $60$ per cent have 3D lengths shorter than $200 h^{-1}$kpc and $59-76$ per cent have 3D distances shorter than $100 h^{-1}$Mpc. Then in \citet{Gimenez-2015} a statistical analysis of the spatial distribution of CGs in the universe was performed, finding that only $27$ per cent of them can be considered to be embedded in larger structures. Also, they mention that $70$ per cent of the embedded groups are 3D physically compact. In \citet{Wiens_2019} the prevalence of 3D CGs was investigated, they use a selection criterion that results in a population of CGs at $z<0.03$ with number densities consistent with observational CGs, also, the great majority of $z=2$ CGs have merged into a single galaxy by $z=0$.

The present work aims to study some properties of CGs populations in different dark matter scenarios. Particularly, we are interested in the characterisation of HCGs arising from N-body simulations corresponding to three different initial seeds associated with specific realisations of the GDM model.

We explored how small variations in the properties of dark matter, in the linear regime, affect the initial seed and thereof the formation of small and medium structures; that decisively determines the CGs counts, the spatial distribution properties and the evolution at different values for the scale factor. We made a complete analysis and comparison of the distributions of different observables corresponding to samples of HCGs for three GDM models and for CDM which were obtained via N-body simulations. Moreover, we confront these distributions against those of an observed HCGs sample, such as the velocity dispersion, number of members, the total mass and the size. 
On the other side, as an intermediate step and to study HCG later, we needed to study the galaxy formation derived from the dark matter halos and their dynamics in the simulations by tracing back the evolution of these halos and constructing the galaxy formation using a semi-empirical model. 
In addition, we implemented an algorithm of classification to identify CG (CGs) of galaxies to explore further these systems. 

The sections are organised as follows, in section \ref{sec:2} we describe the free parameters of the GDM model and their effects on the decay scale. Then in section \ref{sec:3} we selected the parameters for $c_{\text{vis}}^2$, $c_{\text{s}}^2$ and $w$ near zero to construct the Matter Power Spectrum. Also, we described how the cosmological simulations were performed as well as the description to create the mock catalogues. At the end of that section, we presented the Halo and Mass functions to compare models at different scales and redshifts. 
In section \ref{sec:4}, there is a brief description of the different classification criteria used in different works. Additionally, we implemented our group finder algorithm to the mock catalogues. After that, in section \ref{sec:5} we analysed the spatial properties of the simulated HCGs considering two different scenarios and analysed the performance of the algorithm of classification. In section \ref{sec:6} a comparison between GDM models, standard CDM and the observations are performed considering the spatial and intrinsic properties. Finally, in section \ref{sec:7} we present the conclusions.


\section{GDM model}
\label{sec:2}
GDM model describes a general fluid that admits a pressure term and a shear different from zero compared to a perfect fluid, where these quantities vanish. Therefore, the energy-momentum tensor has these two extra parameters 
\begin{equation}
 T_{\mu\nu}=(\rho+P)u_{\mu}u_{\nu}+Pg_{\mu\nu}+\Sigma_{\mu\nu},
\end{equation}
such additional perturbation are controlled by the GDM parameters. Namely, the equation of state relates the background pressure and energy densities $w=\bar{\rho}/\bar{P}$, while $c_{\text{s}}^2$ and $c_{\text{vis}}^2$ control the pressure $\Pi$ and the scalar anisotropic stress $\Sigma$ perturbations, respectively. These last two quantities are expressed through some closure equations \citep{kopp}
\begin{eqnarray}\label{eq: close 1}
\Pi_g&=&c_a^2\delta+(c_{\text{s}}^2+c_a^2)\hat{\Delta}_g\\\label{eq: close 2}
\dot{\Sigma}_g&=&-3H\Sigma_g+\frac{4}{1+w}c_{\text{vis}}^2\hat{\Theta}_g,
\end{eqnarray}
where $\hat{\Delta}$ and $\hat{\Theta}$ are gauge invariant density and velocity perturbations for GDM. In fact, the label $g$ refers to the GDM fluid.
Thus, it becomes clear how the GDM free functions are defined in physical terms.

The structure formation in a GDM universe depends on the values of the free functions due to the magnitude of $c_{\text{vis}}^2$ and $c_{\text{s}}^2$ give rise to the decay of the gravitational potential at scales below that
\begin{equation}\label{eq: decay}
 k^{-1}_{\text{dec}}(\eta)\sim \eta\sqrt{c_{\text{s}}^2+\frac{8}{15}c_{\text{vis}}^2},
\end{equation}
as long as the free parameters are non-zero \citep{kopp}. If $c_{\text{s}}^2\gg c_{\text{vis}}^2$ then the potential starts to oscillate below the Jeans scale. On the other hand, $c_{\text{vis}}^2$ damps the density perturbation without causing any oscillations \citep{thomas_2020}.

The values for these free parameters determine the dynamics of the perturbations and therefore the shape for the Matter Power Spectrum \citep{kopp}. In \citet{thomas_2020}, an analysis using a halo model was made to constrain the values for these free parameters using large scale structure data from various surveys, that is, Planck Power Spectra (PPS), the Baryon Oscillation Spectroscopic Survey Sloan Digital Sky Survey (BAO) and the Planck CMB lensing likelihood.

\section{Cosmological simulations}
\label{sec:3}
\subsection{Matter Power Spectrum}
\label{sec:3a}
In this work we have used values for $c_{\text{vis}}^2$ , $c_{\text{s}}^2$ and $w$ close to those for CDM. We are considering three different scenarios for GDM, namely
\begin{itemize}
 \item \textbf{GDM I}: $w=-1\times 10^{-6}$, $c_{\text{s}}^2=1\times 10^{-7}$ and $c_\text{vis}^2=1\times 10^{-7}$,
 \item \textbf{GDM II}: $w=-1\times 10^{-6}$, $c_{\text{s}}^2=1\times 10^{-7}$ and $c_\text{vis}^2=1\times 10^{-6}$,
 \item \textbf{GDM III}: $w=6\times 10^{-4}$, $c_{\text{s}}^2=1.92\times 10^{-6}$ and $c_\text{vis}^2=1.1\times 10^{-7}$.
\end{itemize}

\begin{figure}
 \centering
 \includegraphics[width=8cm]{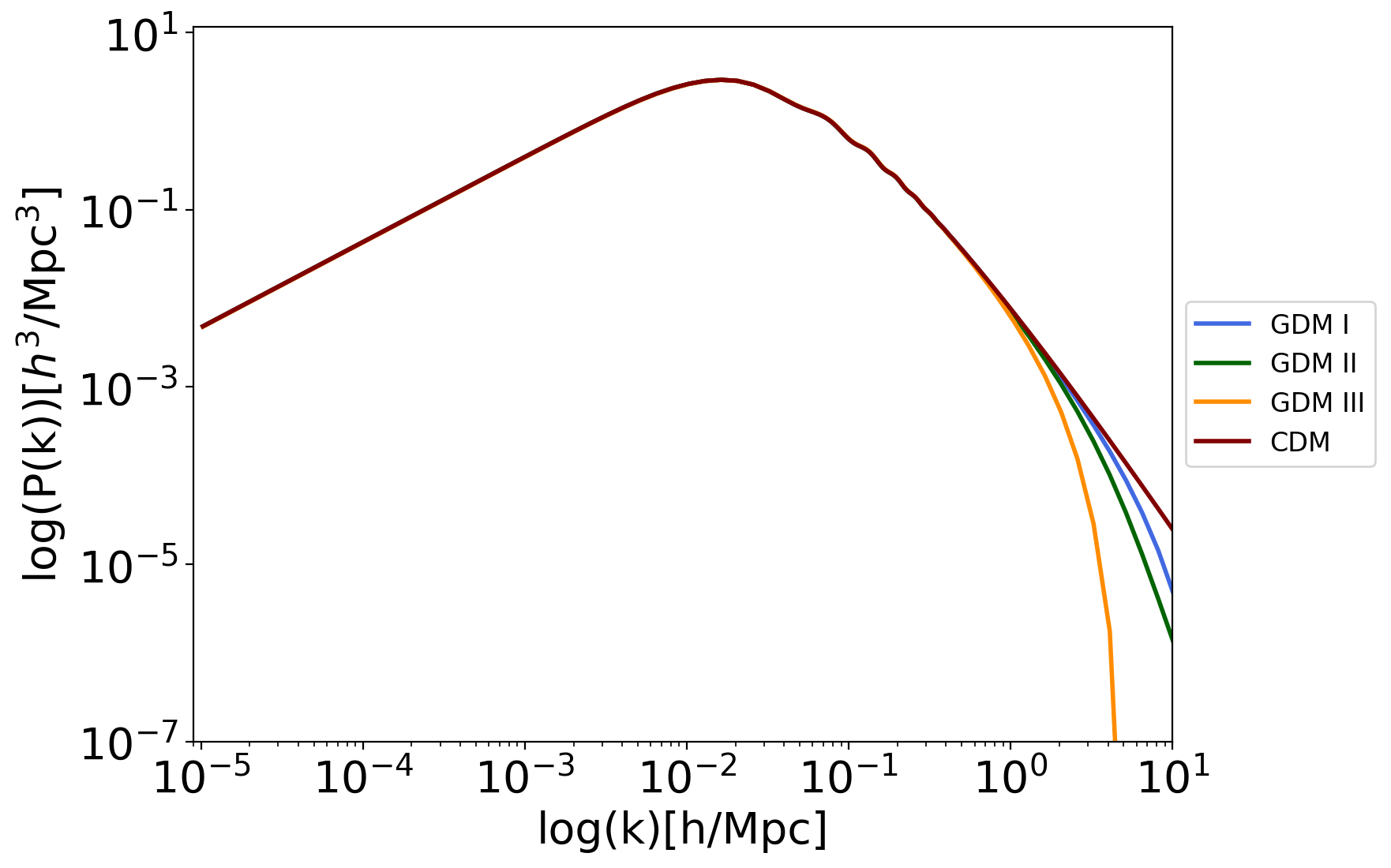}
 \caption{Matter Power Spectra for CDM and three different scenarios for GDM at z=127. Results are identical until $k=1$, from this value, the GDM models start to show a cut-off in accordance with the $c_{\text{vis}}^2$, $c_{\text{s}}^2$ and $w$ parameters, remaining in scenarios near CDM.}
 \label{fig: power_spectrum}
\end{figure}

These values for the parameters are based on the constraints reported in \citet{thomas_2020} mentioned in the last section.
The Matter Power Spectrum was generated by solving the linear perturbation equations using the Boltzmann code GDM-Class\footnote{https://github.com/s-ilic/gdm}\citep{GDMCLASS} which is a modified version of the code CLASS\footnote{https://lesgourg.github.io/class\_public/class.html} \citep{CLASS}. The Matter Power Spectra obtained for the different models are plotted in Fig. \ref{fig: power_spectrum}. In that figure is clearly shown that GDM I is less viscous than GDM II which at the same time is \textit{colder} (it holds smaller $c_{\text{s}}$) and more viscous than GDM III. In the last model, the potential decays earlier according to \eqref{eq: decay}, causing a sharper cut-off. 
Although the differences between GDM I and GDM II parameters may seem minimal, at scales corresponding to $k=1$, GDM I shows a less pronounced cut-off than the subsequent model.
In all the cases, the differences become visible for wavenumber values $k\geq 1$. Below that value of k, corresponding to large scales, the Power Spectrum for the three models are identical. 

\subsection{Cosmological parameters and Initial Conditions}
\label{sec:3b}
To create initial conditions for our simulations we used the N-GenIC code \footnote{https://www.h-its.org/2014/11/05/ngenic-code/} \citep{NGENICa,NGENICb} which constructs an initial distribution of N-body particles from the distribution of density perturbations prescribed by a given matter power spectrum.
The cosmological dm-only simulations were performed by using Gadget 2\footnote{https://wwwmpa.mpa-garching.mpg.de/gadget/} \citep{Gadget2} which is a Smoothed Particle Hydrodynamics (SPH) code that uses a hierarchical tree algorithm to compute the gravitational forces. 
Given that the sound speed and viscosity in our GDM fluid models are very small, we carry out simulations within an approximation where the effects of collisions between the N-body particles at small scales are neglected. Basically, we are interested in studying the effects produced by the cut-off of the initial matter power spectrum (mpk) (due to the free-streaming of linear perturbations) over the non-linear structure formation. 

\begin{figure*}
\centering
\begin{tabular}{cc}
\includegraphics[width=8cm]{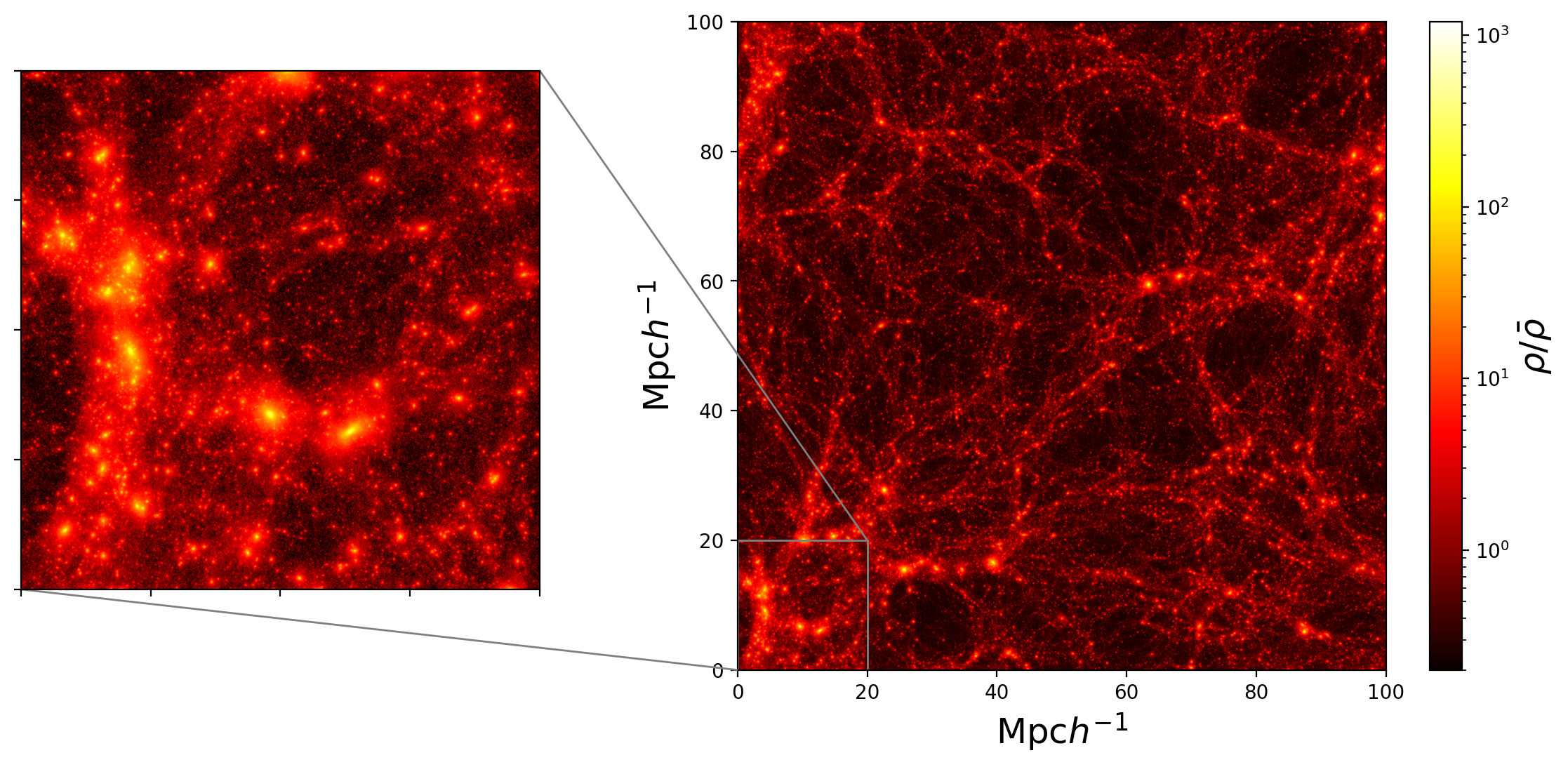}&\includegraphics[width=8cm]{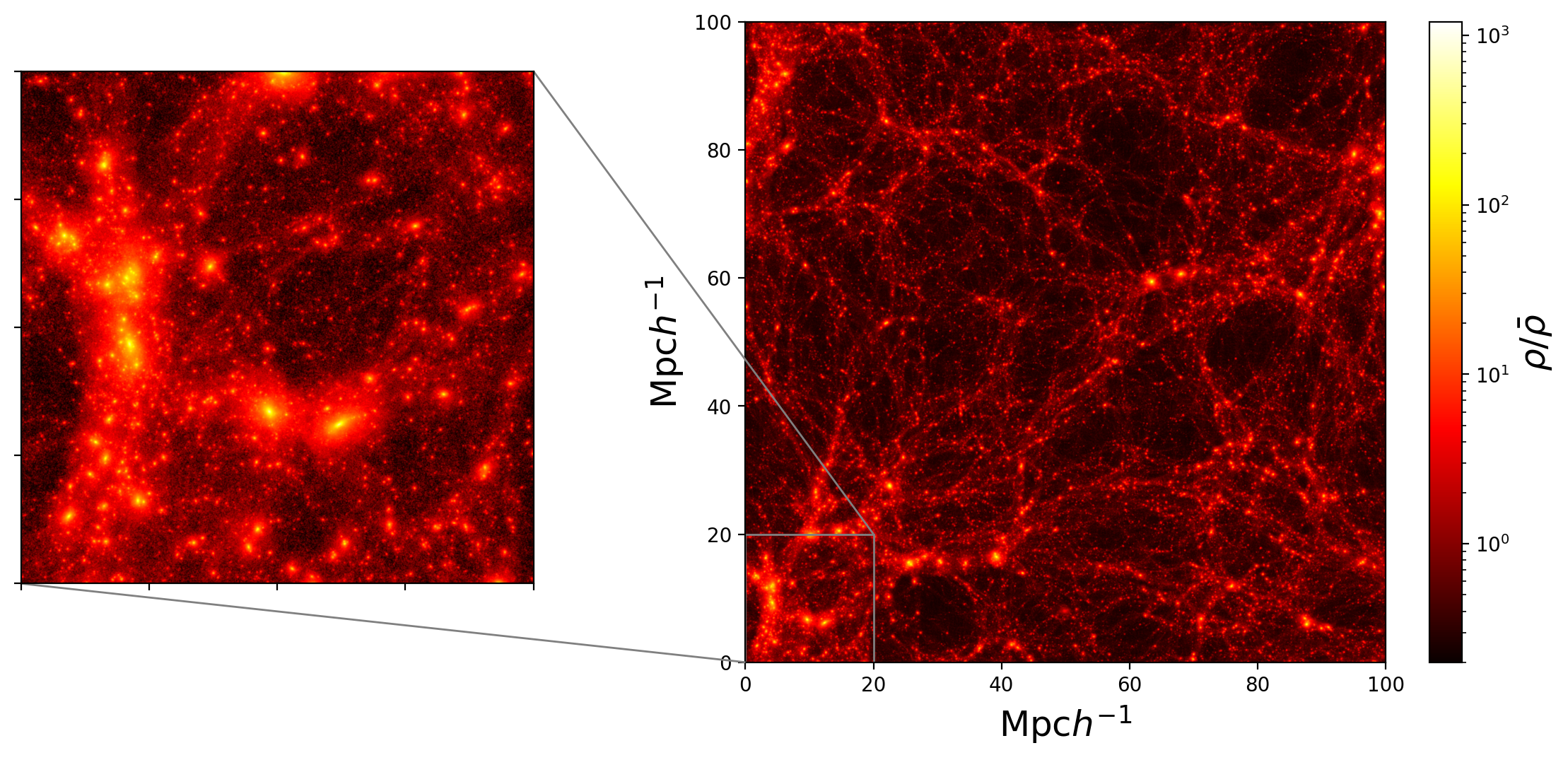}\\
\textbf{(a)}&
\textbf{(b)} \\
\includegraphics[width=8cm]{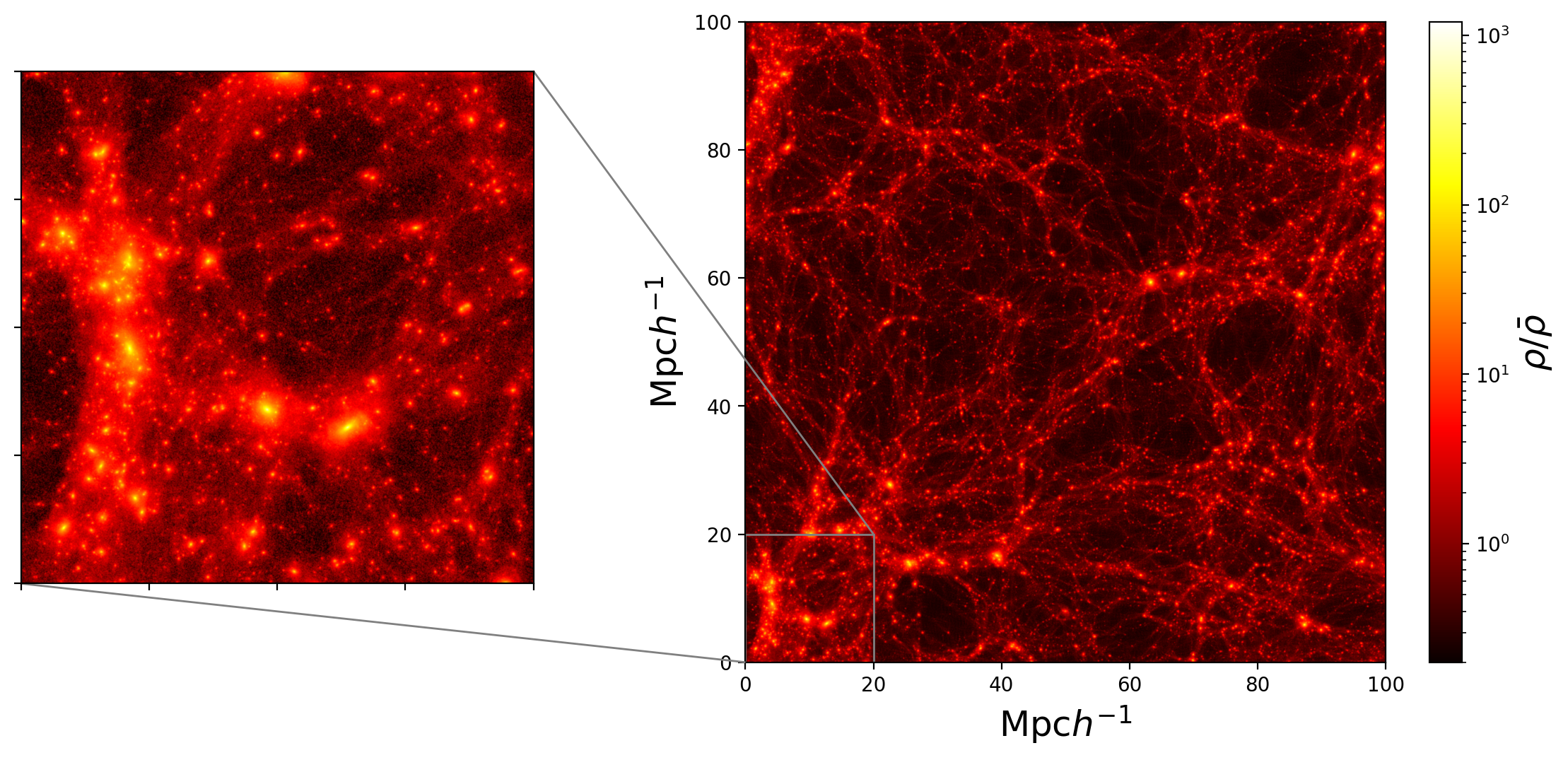}&\includegraphics[width=8cm]{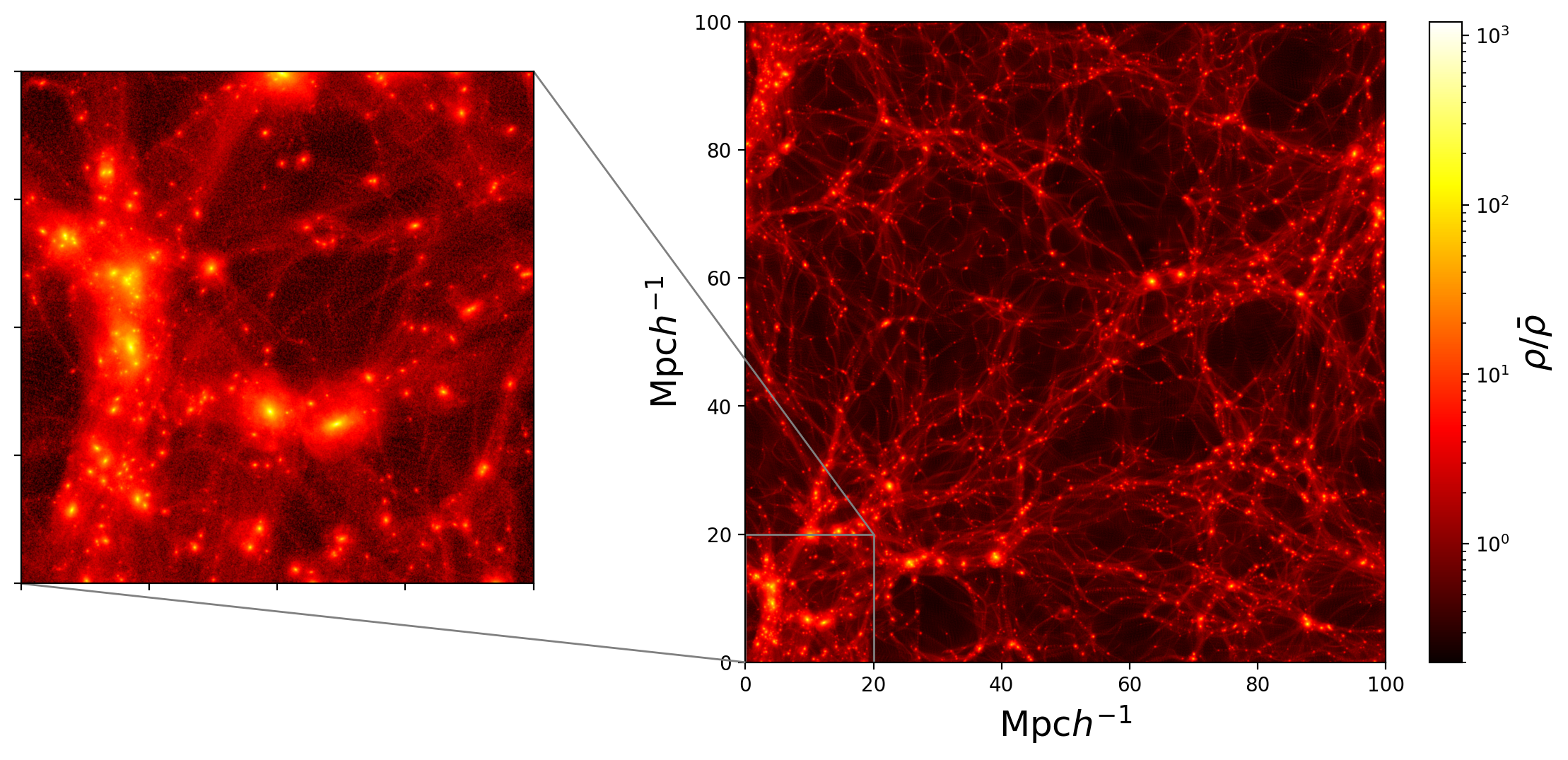}\\
\textbf{(c)} &\textbf{(d)} 
\end{tabular}
\caption{Projections in the XY plane of the simulations for CDM (a), GDM I (b), GDM II (c) and GDM III (d) for $Z\in[0,50]\text{Mpc}/h$. In the left panels, we show an amplified region to observe the differences between models at small scales. For large scales, the structure formation seems very similar. In the high-density regions is clear how CDM has more substructure than the other models. GDM III is the model with the least structure formation and also the voids (i.e. the regions where the contrast has a negative value) are more evident.}
\label{fig: simulations}
\end{figure*}

In all our simulations, the boxsize is $L=100$ $\text{Mpc}/h$ and the particle number is $N_{\text{tot}}=512^3$. The initial redshift is $z=127$. The cosmological parameters are $\Omega_{\text{dm}}=0.25$, $\Omega_b=0.05$, $\Omega_{\Lambda}=0.7$ and $H_0=73 \text{km/s Mpc}$. The $\sigma_8$ value varies in each model, in the case of CDM is 0.84, for GDM I is equal to 0.81, for GDM II is 0.75 and GDM III is only 0.65. The softening length corresponds to $2$ per cent of $(V/N_{\text{tot}})^{1/3}$, where $V$ is the volume of the box, $V=L^3$. 
Snapshots for our simulations at $z=0$ projected onto the $xy$ plane are shown Fig. \ref{fig: simulations} where we used the library Pylians3\footnote{https://pylians3.readthedocs.io/en/master/index.html} for making these plots. As expected, for the GDM instances, an evident lack of structure at small scales can be appreciated as an effect of the power spectrum cut-off.

\subsection{Mock catalogues of galaxies}
\label{sec:3c}

In this section, we aim to study some properties of galaxy populations, therefore, we need information about the dynamics and distribution of the galactic baryonic components. 
Since our cosmological simulations are dark matter-only, in order to introduce information about baryons in galaxies we used a semi-empirical approach as it is usual on other simulations as in \citet{NGENICa, Bolshoi, Horizon,Multidark}. 
By using this information about structure formation provided by the simulations, it is possible to trace back the history of each halo, their progenitors and descendants halos and to identify the successive mergers occurring between them. This reconstruction is known as a {\it merger tree} and it can be linked to the galaxy formation by means of a semi-analytical or semi-empirical model, where the baryonic components within halos are modelled through the physical or observational prescriptions \citep{Delucia-Review}. In particular, semi-empirical models trace the co-evolution of galaxies and their host dark matter halos over time, constraining the galaxy–halo connection at each epoch through the mass accretion and star formation rates.
\begin{figure}
 \centering
 \includegraphics[width=8cm]{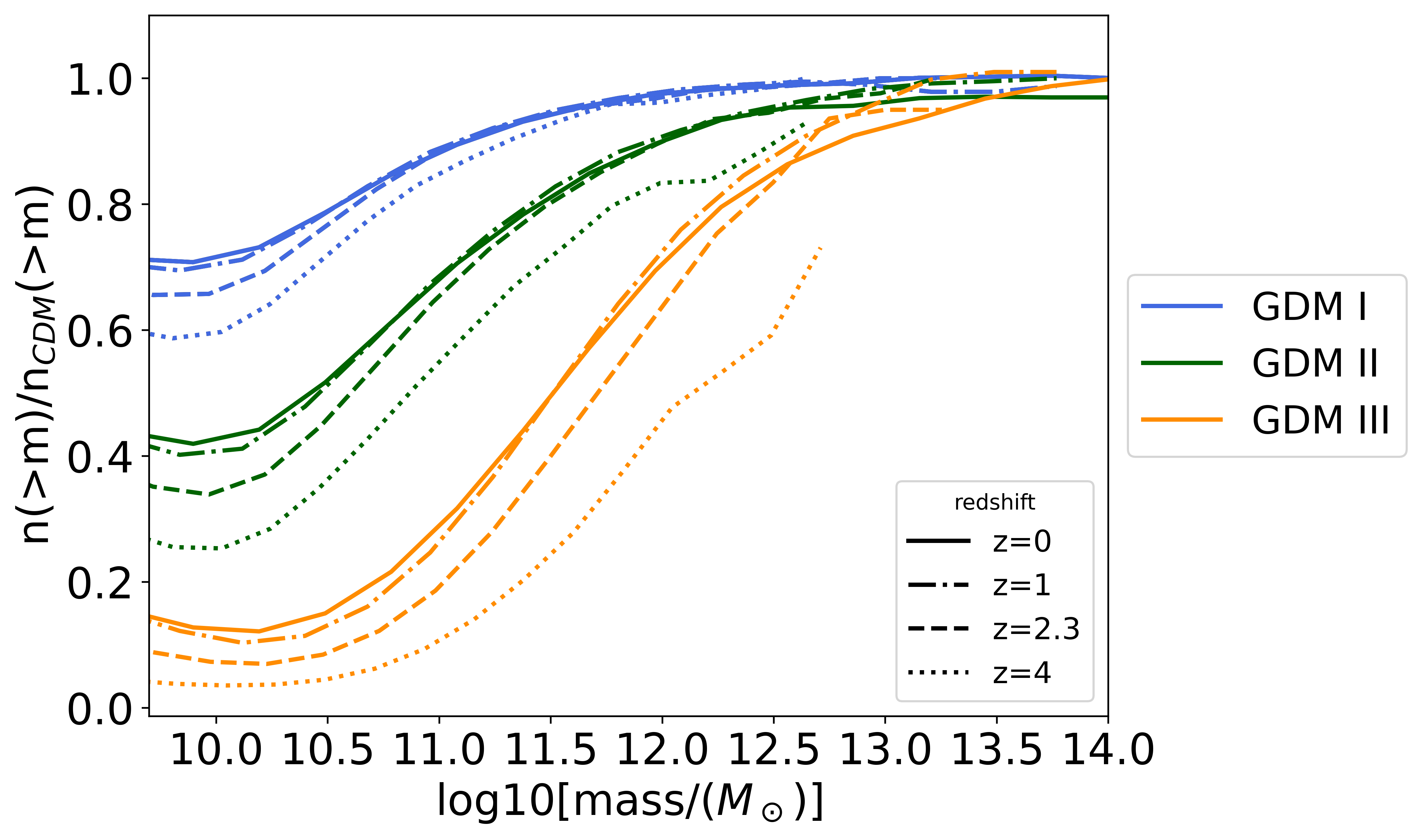}
 \caption{The smoothed relative change for the HMF function associated with different GDM models with respect to CDM at different redshifts. The range of the plot lies between the low limit resolution in the simulations, close to $10^{9}$ $\text{M}_{\odot}$ and the upper limit $10^{14}$ $\text{M}_{\odot}$. The HMF for GDM models show a cut-off in agreement with Fig. \ref{fig: power_spectrum}. For higher z, the large structure is suppressed identically in all the models. Additionally, intermediate structures are less suppressed for models with smaller $c_{\text{s}}$, compared to those at $z=0$. We have changed this plot, in the response we explain why. }
 \label{fig: mass_function}
\end{figure}

In order to identify halos in the simulations, we used the Rockstar finder \footnote{https://bitbucket.org/gfcstanford/rockstar} \citep{Rockstar} which locates dark matter overdensities. Also, we used the Consistent-trees code\footnote{https://bitbucket.org/pbehroozi/consistent-trees} \citep{Ctrees} to generate the merger trees and the Universemachine\footnote{https://bitbucket.org/pbehroozi/universemachine} code \citep{Umachine} to create the corresponding semi-empirical model to describe baryons.

\begin{figure}
 \centering
 \begin{tabular}{c}
 \includegraphics[width=8cm]{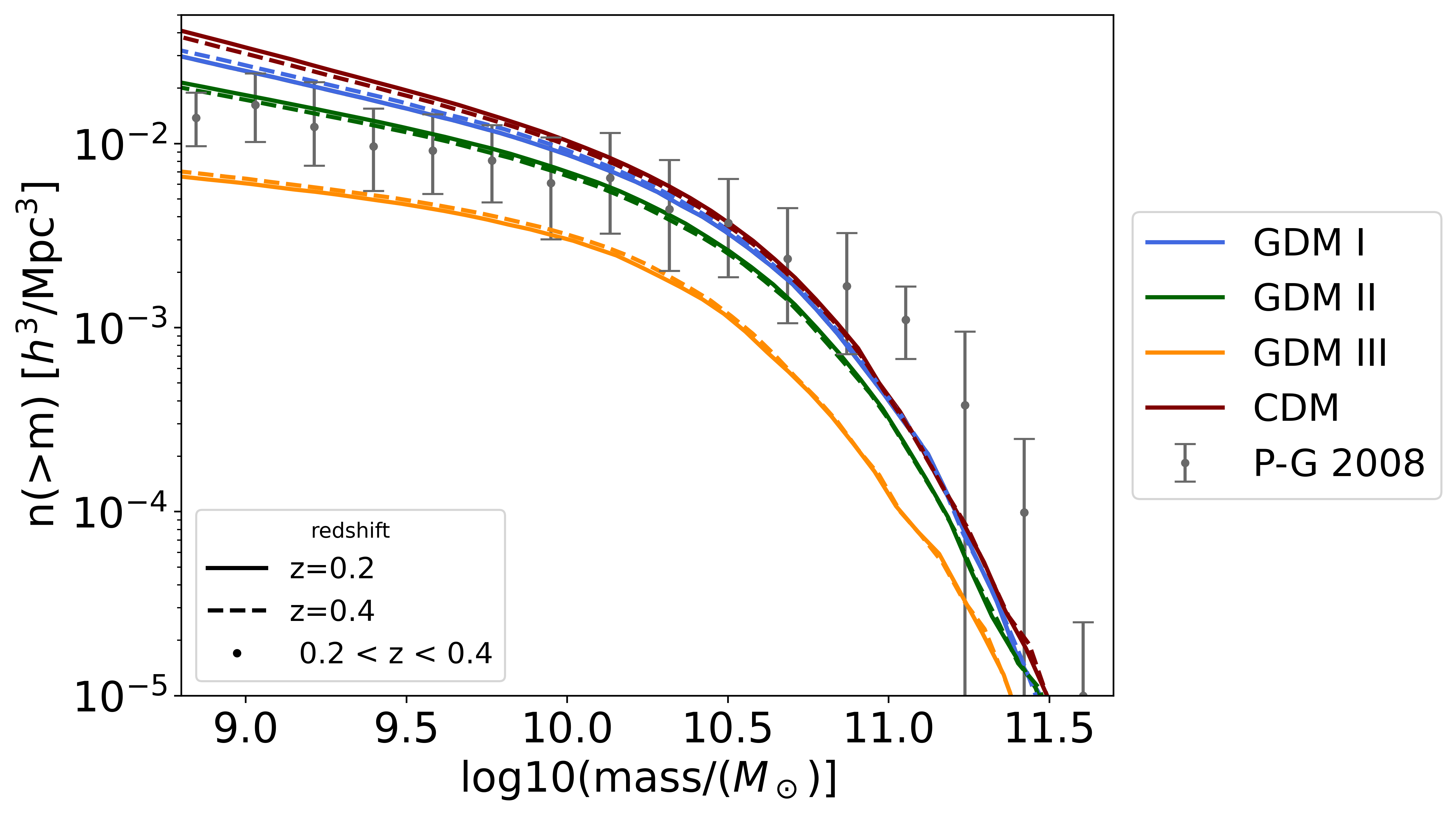}\\
 \textbf{(a)}\\
 \includegraphics[width=8cm]{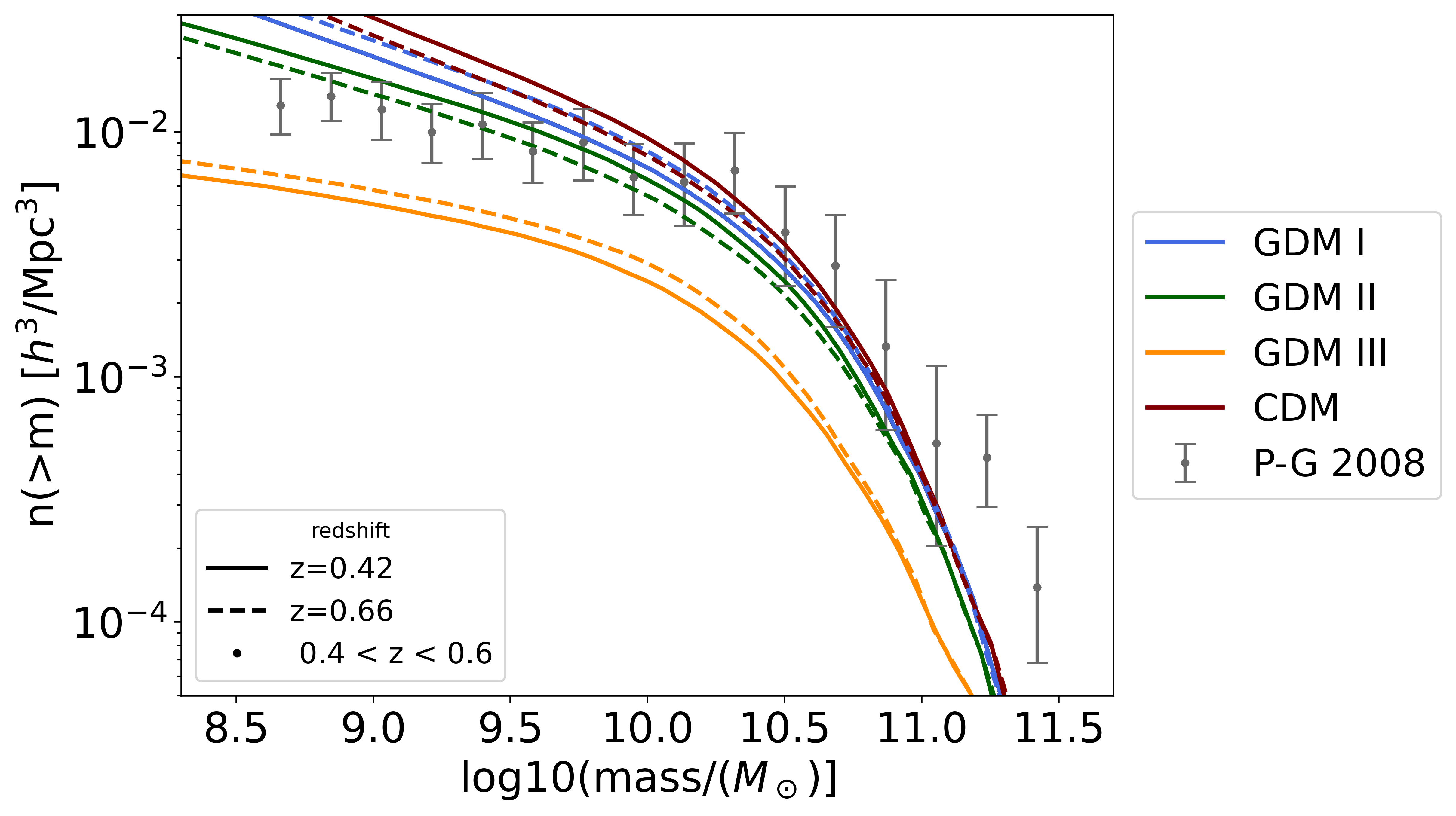}\\
 \textbf{(b)}\\
 \includegraphics[width=8cm]{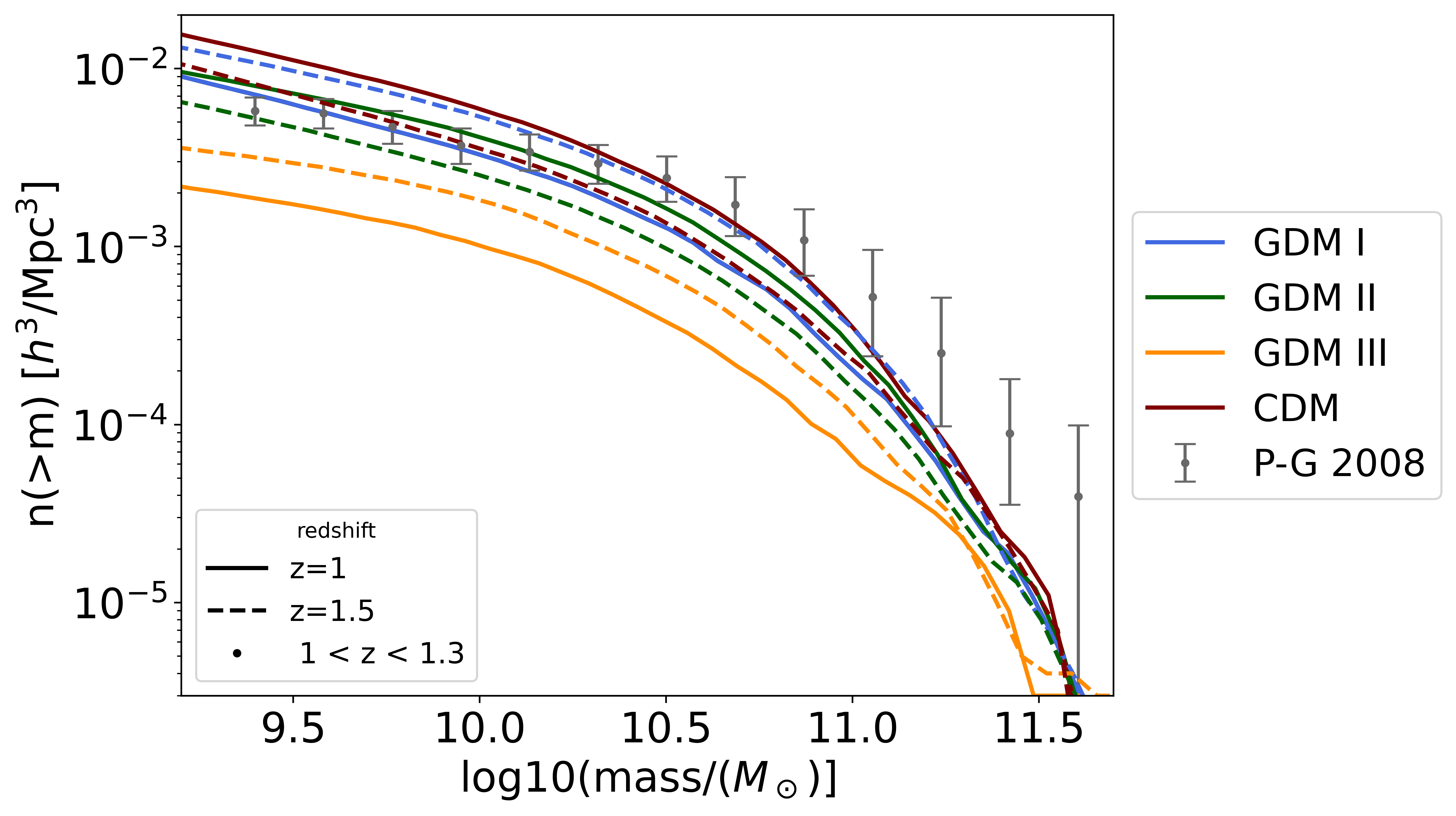}\\
 \textbf{(c)}
 \end{tabular}
 \caption{SMF for the three GDM scenarios and the standard CDM at different redshifts in the range $(0,2)$. It is also shown the SMF for an assembly of galaxies observed by Spitzer which was reported in \citet{Perez-Gonzalez}.}
 \label{fig: smf}
\end{figure} 

\subsection{Halo and Stellar Mass Function}
\label{sec:3d}
In order to quantify the amount of dark matter structures at different redshifts and compare it against the prediction of CDM, we computed the halo mass functions (HMF) in different models for different times relative to that for CDM. The results are shown in Fig. \ref{fig: mass_function}.
Clearly, the mpk cut-off for GDM I, II and III is in agreement with that in\ref{fig: mass_function}. 
Counts of large structures ($M>10^{12.5}M_{\odot}$) within GDM models are similar to those for CDM at $z=0$. When considering higher redshift values, we observe that the number of halos from small to medium scales decrease as $z$ increases. However, counts of the biggest structures show the largest discrepancy given that the most massive halos are not created until redshifts close to zero. 

Additionally, by following the prescription described above, we reconstructed the stellar mass functions (SMF) of galaxies for simulations corresponding to different models at different redshifts. They are shown in Fig. \ref{fig: smf} along with the observational stellar mass functions reported in Table 1 by \citet{Perez-Gonzalez}. In that work, the authors analyse measurements of the stellar mass of a sample of galaxies observed with the Spitzer telescope with redshifts ranging between $z=0$ and $z=4$.

The stellar mass functions for the models described in this work have been plotted at the nearest redshift available in our simulations snapshot to the mean one reported in the observations. Besides, the functions are shown in the ranges of masses reported for the observed galaxies. As we can see in the figure \ref{fig: smf}, the GDM III prediction of SMF disagrees the most from observations for all the redshift values. In addition, the CDM prediction for the density number of galaxies at scales below $10^{10}M_{\odot}$ is overestimated for all redshifts. In contrast, the GDM III prediction of the density number is underestimated for the entire range of masses in all redshifts.
In some cases, the number density of the largest observed stellar structures is largely underestimated. This is also the range where the error bars from the observations increase significantly.

\section{HCG of galaxies}
\label{sec:4}

Since the main focus of this work is the study of properties of HCG in simulations associated with different GDM fluid models. We dedicate this section to mention the basics about them and how they have been usually classified over time.
 
In 1982, Paul Hickson classified a sample of 100 HCGs following the list of criteria:

\begin{itemize}
 \item Population: $\label{eq:popu} N_{\text{pop}}\geq 4$,
 \item Isolation: $\label{eq:islat} \theta_N\leq 3\theta_G$,
 \item Compactness: $\bar{\mu}_G<26$,
\end{itemize}
where $N$ is the total number of galaxies within 3 mag below that of the brightest member, this criterion allows selecting galaxies with similar masses instead of very massive members and their satellites. Also, $\bar{\mu}_G$ is the average total magnitude of these galaxies per arcsecond$^2$ inside the smallest circle (with angular diameter given by $\theta_G$) that contains their geometric centres. Finally, $\theta_N$ is the angular diameter of the largest concentric circle containing no other (external) galaxies within this magnitude range \citep{Paul1982}.

In some works, such as \citet{Barton} or in \citet{Tzanavaris}, triplets with $N_{\text{pop}}\geq 3$ are taken into account; for instance, the Redshift Survey (RSCG’s). In this work, we follow the standard selection criterion considering $N=4$ as the minimum number of members, Regarding point 4: given that we consider observations of HCGs from the original catalogue which was constructed using this criterion. However, in the first part of the analysis we also consider triplets in order to study some effects arisen from modifying the classification criteria.

Also, in other works additional considerations are taken into account as in \citet{Gimenez-Mamon}, aiming to avoid selection biases, the following condition is imposed besides the standard ones:
$$
 R_{\text{brightest}}\leq 14.44 (\text{flux limit}),
$$
where $R_{\text{brightest}}$ is the R-band magnitude of the brightest galaxy. 
Additionally, some restrictions to the members velocity are placed such as: 
$$|v_i-\tilde{v}|\leq 1000 \text{ km s}^{-1},$$
where $v_i$ is the radial velocity of the i-th member and $\overline{v}$ is the median radial velocity of the set \citep{Gimenez-Mamon,Wiens_2019,Gimenez-2015}. This last works as a filter of \textit{fly-by} galaxies which are not actual members, that is, they are not bound by the group. It is important to mention that this filter was applied in subsequent works by Hickson for the observational HCGs \citep{hickson-1992}.

On the other hand, although in literature different algorithms using 2D projection of data to identify HCG in N-body simulation have been proposed \citep{McConnachie,Gimenez-Mamon,Gimenez-2015,Farhang} (aiming to mimic the selection procedure applied to observations), it is possible to establish a relation between the projected separation and the 3D distance between members \citep{Wiens_2019} and therefore, to modify the original compactness criteria according to the spatial information (which is available in the mock catalogues). 
For that reason, it is useful to take as a reference the reported median projected separation between members $39 \text{kpc}/h$ inferred from observations, which interestingly is comparable to the sizes of the galaxies, and the median redshift of the sample of HCG $z=0.0297$. It is important to mention, that the last value strongly depends on the observational capability achieved up to date. In this work, these observational reference values are used to determine the relation between 3D and projected distances.

\subsection{Group finder algorithm for mock catalogues}
\label{sec:4a}

We adapted the selection criteria described above to the 3D available information in the mock catalogues as follows:

\begin{itemize}

 \item[1)] \textbf{Compactness} Instead of defining the compactness of a HCG by using the surface brightness of the group as it is done in the original selection rule, we consider the observational reference value of the median projected separation between members. This criterion will be applied once the relation between the median projected separation and the median 3D separation for the simulated HCGs candidates is established (Section \ref{sec:4b}). 
 
 \item[2)] \textbf{Isolation}: We constructed a shell with size three times the radius of the HCGs candidates ($r_\text{shell} = 3r_{\text{group}}$). Then, the ratio of the density inside that shell to the density of the group is computed. A candidate is isolated as long \textbf{as} this ratio is sufficiently close to zero, i.e.
 \begin{equation}
 \frac{\rho_{\text{shell}}}{\rho_{\text{group}}}<10^{-4}.
 \end{equation}
 \begin{itemize}
 \item[2a)] \textbf{Boundary:}
 Given that simulations are done within a finite box, clusters located close enough to the boundary are dropped; namely, clusters obeying $x_i-r_\text{shell}<0$ or $x_i+r_\text{shell}>$L, where L is the box size and $x_i$ are the Cartesian coordinates of the HCG candidate. These clusters close to the boundary are ruled out since they would fulfil the isolation criterion because a portion of the shell is outside the boundary.
 \end{itemize}
 
 \item[3)] \textbf{Dwarf mass limit} We establish a lower bound for the galaxy mass given by $M_{\text{dwarf}}=2\times 10^{9}M_{\odot}$, in order to get rid of dwarf galaxies \citep{Wiens_2019}.
 
 \item[4)] \textbf{Membership condition}
 We applied the filter over the velocity of members
$
 |v-\bar{v}|<1000 \text{km}/s,
$
 to avoid the \textit{fly-by} galaxies.
 \item[5)] \textbf{Galaxy mass ratio} By assuming that the dominant member galaxy in luminosity corresponds to the dominant galaxy in mass \citep{Wiens_2019}, we are able to select groups of galaxies with similar large masses instead of satellite galaxies. Therefore, HCGs candidates must satisfy the following condition
 \begin{equation}
 \frac{M_2+M_3}{M_1}>0.1,
 \end{equation}
 where $M_1$, $M_2$ and $M_3$ stands for the first, the second and the third most massive galaxies, respectively.

 \item[6)] \textbf{Minimum number of members} After applying the previous filters, we double-check the total number of galaxies in each group. We are considering two instances, the one adopted in original Hickson selection criteria, $N\geq 4$, and the one including galaxy triplets, $N\geq 3$ (see Section \ref{sec:4}).
\end{itemize}

\subsection{Implementation of the algorithm of classification}
\label{sec:4b}

We have used DBSCAN \citep{dbscan}, a tool to identify clusters of data by locating core samples of high density. In our case, it served to classify clusters of galaxies. DBSCAN uses two parameters, $\epsilon$ corresponding to the maximum distance between two neighbour samples and the minimum number of neighbours $N$ required to consider a set of neighbours a cluster. For each simulation we explored the following range of parameters for $\epsilon$, $[10, 120] \text{ kpc}/h$, and $N=[3,4]$ at z=0.

Once the overdensities have been located by the DBSCAN algorithm, selection criteria from 2) to 5) listed above are applied. Up to this point, the compactness indicator 1) is left unspecified until later.

\begin{figure}
 \centering
 \includegraphics[width=8cm]{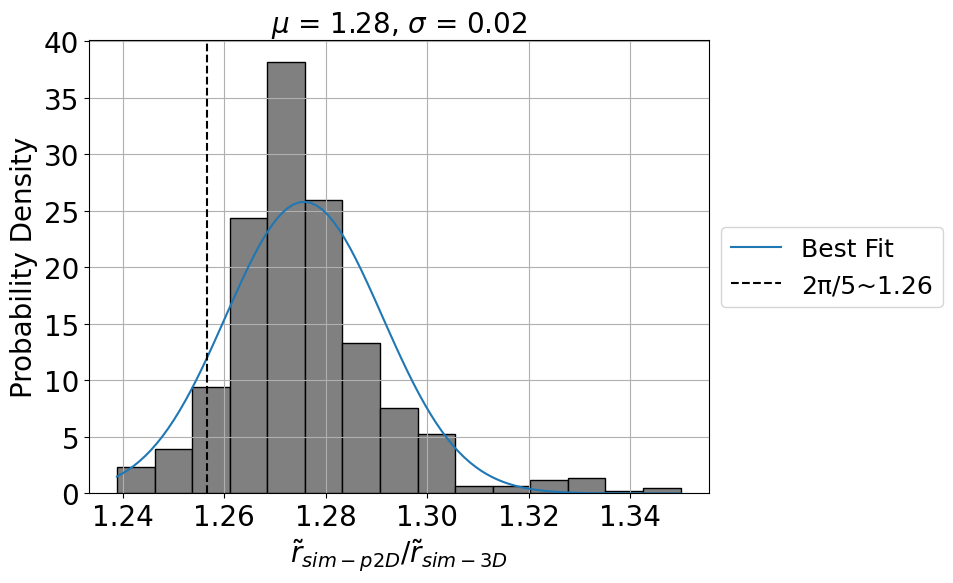}
 \caption{The ratio of the median separation in 3D and median projected separation for all the simulations using the HCGs candidates found for several $\epsilon \in [10,120] \text{ kpc}/h$, considering only results with at least 10 HCGs candidates catalogued per simulation. The dashed line stands for the value obtained in \citet{Wiens_2019}.}
 \label{fig:histMuMup}
\end{figure}

In order to compare with observational reference values, we computed the median projected separation $\tilde{r}_{\text{sim-p2D}}$ between members in the HCGs candidates for $N=4$ by projecting the simulated galaxies in each face of the simulation box. Then, we calculate the physical median distance $\tilde{r}_{\text{sim-3D}}$ between those members and compute the distances ratio $\tilde{r}_{\text{sim-p2D}}/\tilde{r}_{\text{sim-3D}}$. Results are shown in Fig. \ref{fig:histMuMup}, where the whole data for all models are taken into account. The mean value is $1.28\pm 0.03$ within a confidence level of 2$\sigma$. Interestingly it is close to the value reported in \citet{Wiens_2019}, where a Monte Carlo simulation was performed to estimate that the 3D separation between galaxies is roughly $2\pi/5\sim 1.26$ times the corresponding projected separation. 

The mean projected separation between members for the observed HCGs is around $39 \text{kpc}/h$, then the median physical separation between members of observational data should lie within $\tilde{r}_{\text{obs-3D}} = 49.78\pm 1.26 \text{ kpc}/h$ according to the result found in \citet{Wiens_2019}.

\begin{figure}
 \centering
 \begin{tabular}{c}
 \includegraphics[width=8cm]{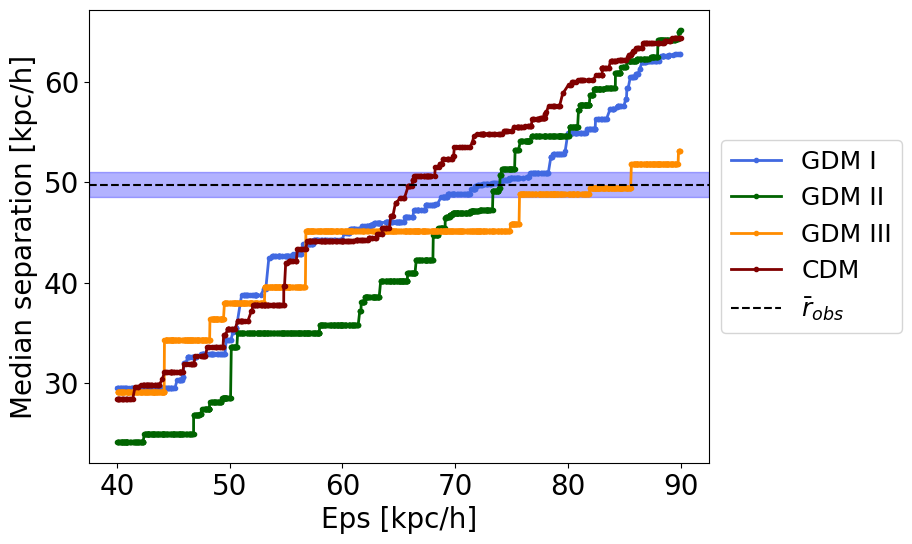}\\
 \textbf{(a)}\\
 \includegraphics[width=8cm]{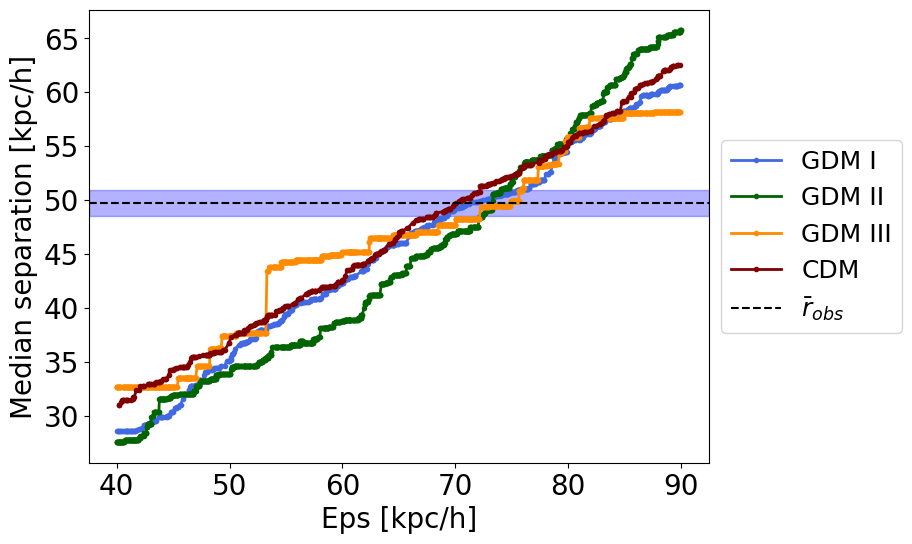}\\
 \textbf{(b)}
 \end{tabular}
 \caption{Result of the DBSCAN and HCGs filter applied to the simulated data for $N_{\text{min}}=4$ (a) and $N_{\text{min}}=3$ (b), observational results are considered as $r_{\text{obs}}=39\pm 2 \text{ kpc}$. The parameter space is plotted in the region near the observational band in both cases.}
 \label{fig: epsvsmpd}
\end{figure}

The relation between $\epsilon$ and 3D median distance of the simulated sample, $\tilde{r}_{\text{sim-3D}}$, is shown in Fig. \ref{fig: epsvsmpd} for both scenarios, N=4 (6a) and N=3 (7b) in different models, where the blue band corresponds to delimited by $\tilde{r}_{\text{obs-3D}}$. Clusters lying within this region are considered to be HCG, this condition defines the compactness criterion. In both cases, the value of $\epsilon$ matching the observational region is different in each model within a range $(65,85) \text{ kpc}/h$.

In addition, the physical separation as a function of $\epsilon$ is increasing in average, however, in some local intervals of the domain, the slope is close to zero. The number of such intervals increases as the GDM parameters are larger, thus this fact is more evident within the GDM III scenario. Also, these intervals are more frequent for $N=4$ and can be related to the fact that the distribution and amount of structure at small scales in each model, for models with less structure the $\epsilon$ value must vary over a large range before presenting a significant change in the median distance. The latter indicates that agglomerations of three galaxies are easier to form than those with $N=4$.

\begin{figure}
 \centering
 \begin{tabular}{c}
 \includegraphics[width=8cm]{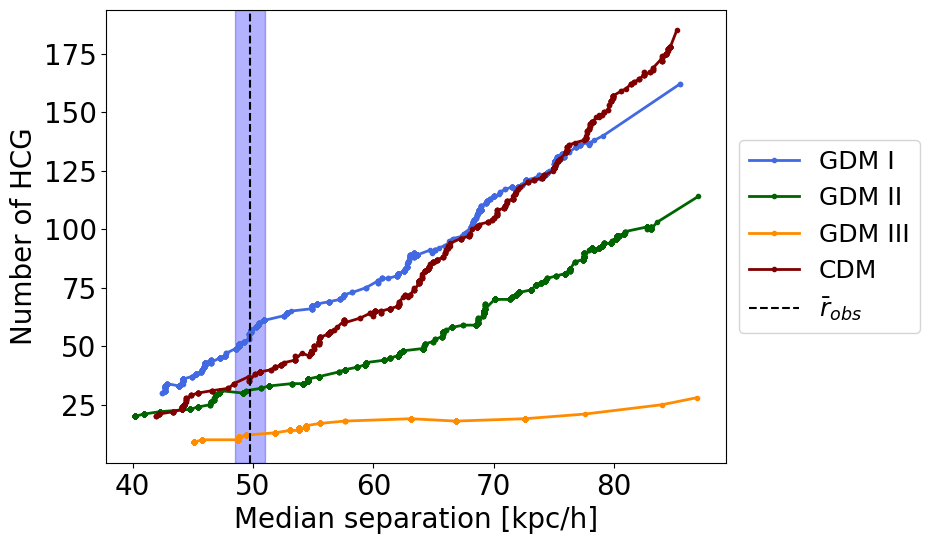}\\
 \textbf{(a)}\\
 \includegraphics[width=8cm]{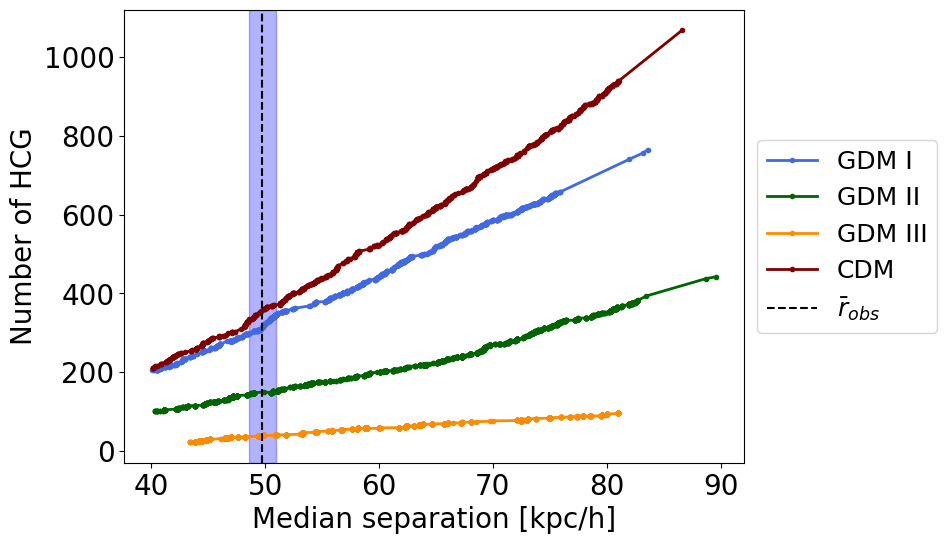}\\
 \textbf{(b)}\\
 \end{tabular}
 \caption{Results of the DBSCAN and HCG filter applied to simulation data. Here we show the physical distance versus the number of HCGs candidates. In (a) for $N=4$, GDM I show a larger number of groups for the median separation laying below $~70 \text{ kpc}/h$, for larger values counts for CDM are dominant. In the same way, the cross-over scale for counts of GDM II and CDM models lay close to the purple band region. In (b) results for $N=3$ are shown, counts for GDM I and CDM are nearby for small values of the $\epsilon$. Here, GDM II and III remain below CDM in all the ranges. }
 \label{fig: epsvsnhcg}
\end{figure}

\section{Spatial properties of simulated HCGs}
\label{sec:5}

We quantified the number of groups found in each model versus the corresponding 3D separation, again, for both cases $N=4$ (a) and $N=3$ (b) in Fig. \ref{fig: epsvsnhcg}.
For $N=4$ clearly the number of HCGs candidates within GDM I is the largest for $r_{\text{sim-3D}}$ ranging within $(40, \sim 74)\text{ kpc}/h$, counts for CDM within that range are slightly smaller, followed by GDM II and GDM III. In contrast, the cross-over scale for GDM II and CDM lines corresponds to $r_{\text{sim-3D}} \sim 43-48 \text{ kpc}/h$. Thus, below this point for GDM II, a larger number of HCGs were found in comparison to CDM count, this behaviour means that GDM models can predict a larger number of HCGs when the substructure in the initial seed is suppressed.
This brings up the question of whether the line of a model with a sharper cut-off in the initial mpk will intersect CDM at a lower value for the median physical separation. If so, there is a balance between low-scale structure suppression and proliferation of HCG.

The situation for $N=3$ is slightly different. First looking at Figure \ref{fig: epsvsnhcg} (b), we can see that it is more likely to find agglomerations of three members. Secondly, counts for CDM are the largest. Even though GDM I is closer to CDM within the range (40,50)\text{ kpc}/h, this one and the rest of the models do not intersect CDM staying far below it.

\subsection{Host halos}
\label{sec:5a}
According to N-body simulations, as some dark matter halos of individual galaxies merge, larger common halos can be created \citep{Barnes-1984,Bode-1993}. 
Actually, dark matter of HCGs is distributed mainly in the intergalactic medium and studies about the kinematics of HCGs environment, indicate that dark matter is mainly placed around the optical galaxies and not inside them \citep{hickson-cg}.
In addition, it has been reported that more than $50$ per cent of the HCGs identified in large N-body simulations are embedded in a common halo \citep{McConnachie}. 

In this work, by considering the mock catalogues for different models, we identified the host halos and their embedded substructure. 
We found that $88$ per cent of the HCGs in CDM are embedded in a host halo and the galaxy associated with that central halo also belongs to the group. The remaining $12$ per cent HCGs are embedded in a host halo, however, its corresponding central halo galaxy is not a member. For the GDM I model, $98$ per cent HCGs have a common halo, while for GDM II and GDM III $96$ and $83$ per cent respectively. In addition, the structure within the host halos is not necessarily part of the HCGs, that is, in some cases, there are other halos and their associated galaxies that belong to the environment rather than the group.
This encourages us to further study the HCGs environment in the future, such as the characteristics of neighbouring structures and their impact on the dynamics and distribution of HCGs and their populations.

\subsection{On the relevance of the different filters when identifying HCGs}
\label{sect:5b}

\begin{figure*}
    \centering
    \includegraphics[width=17cm]{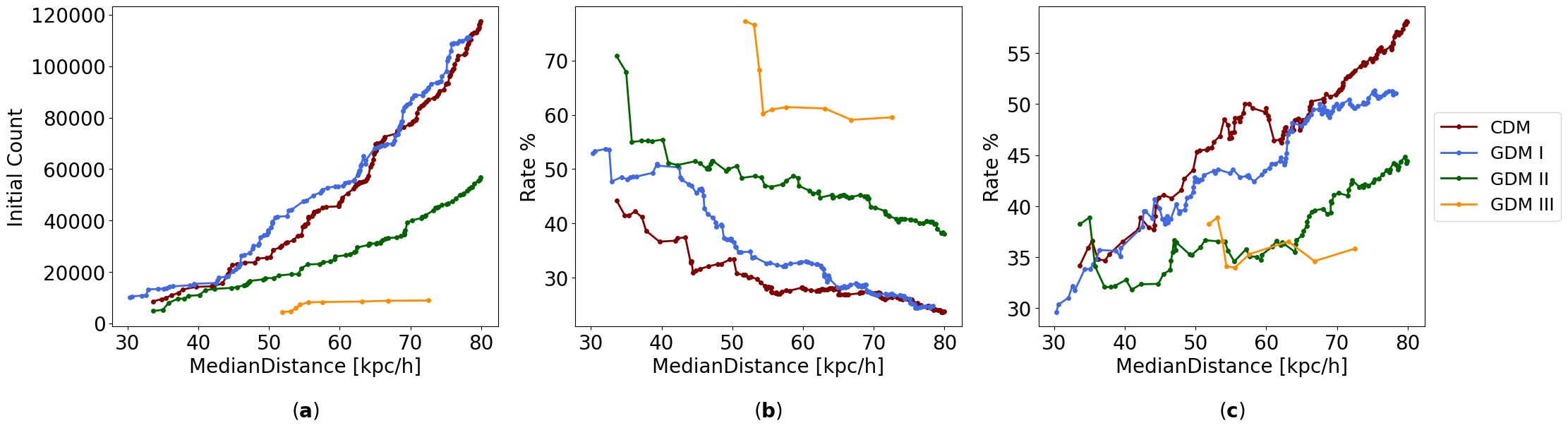}
    \includegraphics[width=17cm]{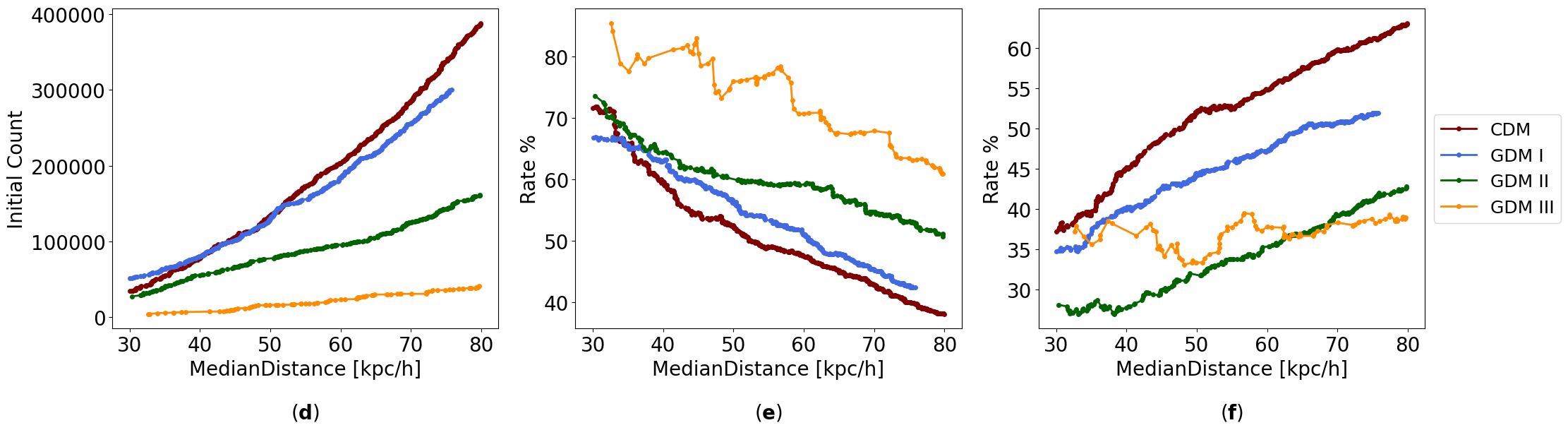}
    \caption{Initial counts of clusters (left), Isolation (center) and mass ratio (right) filter survival rate of clusters for n=4 (above) and n=3 (bellow). Only data with at least 12 HCG remaining after applying all filters were considered.}
    \label{fig:rates}
\end{figure*}   

In Fig. \ref{fig:rates} we present the initial count of clusters classified by DBSCAN as well as the survival percentage of clusters as a function of the mean members separation after the principal filters are applied, namely, isolation and mass-ratio filters either for $N=4$ (upper panels) and $N=3$ (bottom panels). Interestingly, for $N=4$ (a), the initial cluster count is slightly higher for GDM I than CDM, which can be explained by the fact that highly-populated-clusters are more likely in CDM while groups within GDMI tend to be less dense. This behaviour is also shown for $N=3$ (d) for mean distances below than $50 \text{ kpc}/h$, and beyond that value CDM initial clusters become more likely. Besides, GDM II and III a small initial number of clusters, being the latter the one with the lowest count, as a consequence of the pronounced deficit of small structures. 

\begin{figure}
 \centering
 \begin{tabular}{c}
 \includegraphics[width=8cm]{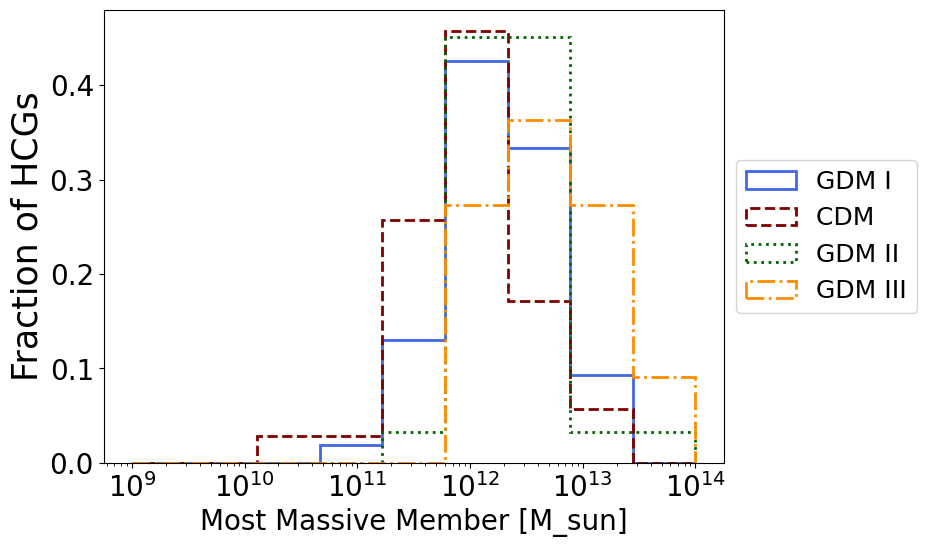}\\
 \textbf{(a)}\\
 \includegraphics[width=8cm]{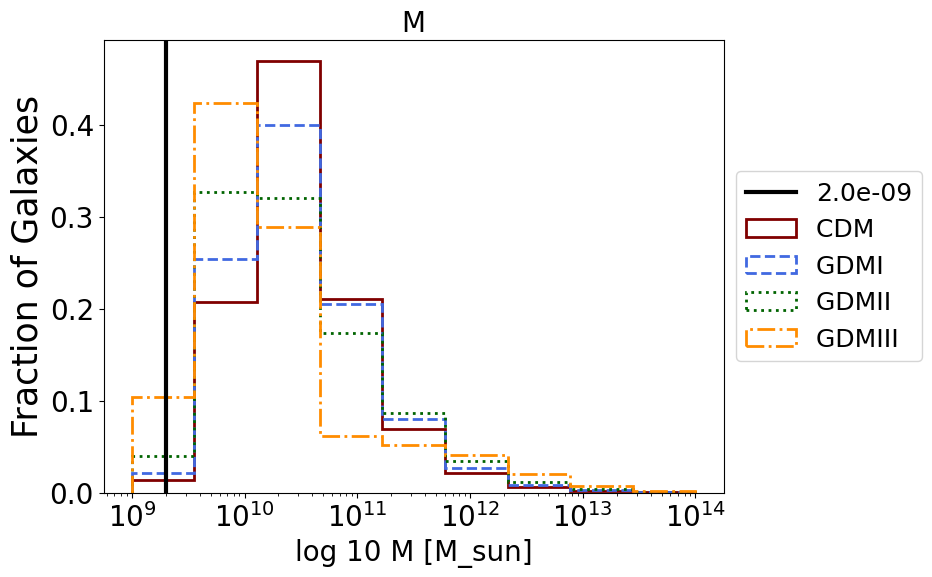}\\
 \textbf{(b)} 
 \end{tabular}

 \caption{(a) Histograms of the mass of the largest HCGs members for the N=4 sample. As expected the distribution of masses in GDM III simulations is shifted to large values since small substructure is seldom compared to CDM, therefore, it is more likely to find large galaxies than small ones. (b) Histograms of the virial masses of galaxy-sized halos within the mock catalogues for different models. The black vertical line corresponds to the dwarf limit.}
 \label{fig: mostmassive}
\end{figure}

In addition, we have computed the survival percentage of clusters after the isolation criterion is applied, again either for $N=4$ (b) and $N=3$ (e). In this case, we can see that counts for CDM are the lowest. The vast amount of substructure within this model prevents fulfilling this condition,  in contrast to what happens within the GDM model, where the percentages are larger.
These differences can be explained by the fact the slopes are different for each model in $N=4$. Besides, for $N=3$ this hierarchical behaviour is also reproduced, namely GDM III is less affected than GDM II, GDM I and of course, CDM.
Additionally, it is worth to mention that
the dwarf limit has a negligible impact on the counts since for the given resolution, galaxies of intermediate masses are more likely to be ruled out. A similar situation arises for the fly-by restriction. Thus, since counts are not sensitive to these last filters, we do not show them in Fig. \ref{fig:rates}.
Also, we have computed the survival percentage of clusters after the mass-ratio condition is applied. According to Fig. \ref{fig:rates} (c) and (f), the more pronounced cut-off in the initial mpk for a given model, the more 
clusters are ruled out by this filter. Additionally, the slopes are similar in all cases for both $N=4$ and $N=3$. 
To get an idea, consider for instance GDM III simulations. Firstly, small structures are the most rare within this scenario, so the probability of forming clusters is smaller than in other scenarios. 
In order to verify the previous assertion, Fig. \ref{fig: mostmassive} can be helpful. Panel a) in that figure shows that, on average, within GDM III and GDM II simulations, the most massive HCGs members masses of $3\times 10^{13} M_\odot$ and $4\times 10^{12} M_\odot$ respectively (see Fig. \ref{fig: mostmassive} b)). In contrast, for CDM, the mean value is $2\times 10^{12} M_\odot$ and there is no significant contribution of galaxies with masses higher than $1\times 10^{13} M_\odot$.
This means that in CDM, galaxies with masses differing by more than one order of magnitude can cluster together owing to their high abundance. In contrast to GDM III where it is more unlikely to cluster galaxies with too different masses. Therefore, the remaining HCGs are made of the most massive galaxies within this last scenario. 
Finally, the last "boundary" filter does not produce any physically relevant effect.  Nevertheless, its purpose is to reduce counting errors in a similar way as the min-halos constraint works. By means of the latter, a double-check of the minimum number of members at the end of the selection procedure is carried out.
An interesting fact observed in Fig. \ref{fig: epsvsnhcg} is the cross-over between CDM and GDM I models for $N=4$, this behaviour can be explained using the Fig. \ref{fig:rates}. On one hand, by considering mean separations below $50 \text{ kpc}/h$, we can observe an excess in GDM I initial number of clusters relative to CDM. In this case, after applying the isolation filter, the percentage of survival clusters is significantly larger for GDM I than for CDM, because of the differences in the slopes of these curves. The effect of the mass-ratio condition on the survival percentage is similar in both models. 
Therefore, the combination of the excess in the initial cluster count and the effect of the isolation criteria explains why the final amount of HCGs is higher within GDM I than in CDM around $50 \text{ kpc}/h$ for $N=4$.
On the other hand, for values of the median separation above the cross value, the initial amount of clusters for the same case ($N=4$) are quite similar, however, given that the isolation survival percentage is practically equal either in GDM I and CDM, the mass ratio filter plays a key role. Since this filter rules out more clusters for GDM I than for CDM, this explains why at these large scales CDM clusters are more abundant than for GDM I. 

In constrast, for $N=3$ the combined effect of both filters produces an excess of CDM counts over GDM I ones as the median separation increases.

\section{Comparison with observations}
\label{sec:6}

In this section, we present a comparison of some metrics describing the geometric and physical properties of the HCG populations either in different theoretical models and observations. In the first place, we compare distributions for HCG in simulations against those in observations. Afterwards, we compare different levels of agreements between models to give a qualitative analysis of the viability of GDM scenarios. 

\begin{figure*}
 \centering
 
 \includegraphics[width=16cm]{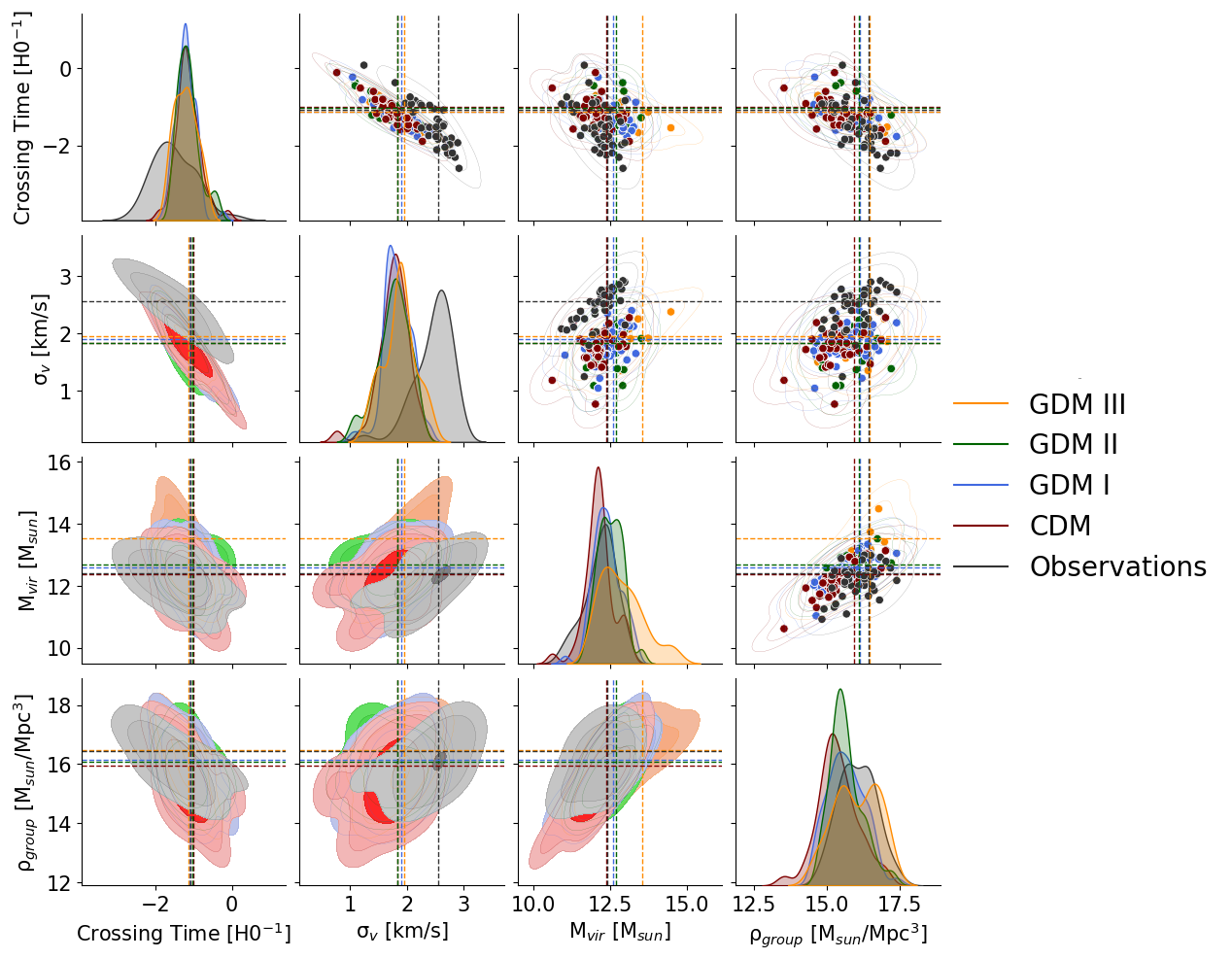}
 \caption{Comparison between observational and simulated data. We present the parameter space between virial mass, density, crossing time and velocity dispersion in a matrix plot, with theirs respective histograms in the diagonal. We used kernel density estimation in the triangular-lower component with 1, 2 and 3-$\sigma$ confidence levels in the contours, whereas in the triangular-upper component we show the raw data. }
 \label{fig: obs-histo}
\end{figure*}

In Fig. \ref{fig: obs-histo} we present the comparison between observed HCGs properties and those for the simulated data in general. Regarding point 4: Due to the fact we are considering the original catalogue data, only the case for $N=4$ will be taken into account in this section and the subsequent.
Fig. \ref{fig: obs-histo} (a) shows 2D distributions in the HCG mass-density ($M_{\text{vir}}-\rho_g$) plane for different models and observations. In there, we observe overlapping regions in all cases. 

Discrepancies between mass distributions between models can be understood in virtue of the properties of the galaxy mass distributions in mock catalogues (as explained in the last section via Fig. \ref{fig: mostmassive}) which result from different clustering mechanisms occurring in different models (see Section \ref{sect:5b}).

The mean value of HCGs mass within CDM, GDM I, II and III simulations are $2\times 10^{12}M_\odot$, $3\times 10^{12}M_\odot$, $4\times 10^{12}M_\odot$ and $3\times 10^{13}M_\odot$ respectively, while the mean value for the observations is $2\times 10^{12}M_\odot$. The statistical values corresponding to the mean, median and mode for the parameters shown in Fig. \ref{fig: obs-histo} are shown in Tables \ref{tab:stat_mass}, \ref{tab:stat_tcr},\ref{tab:stat_veldisp} y \ref{tab:stat_density}. 

Discrepancies between mass distributions between models can be understood in virtue of the properties of the galaxy mass distributions in mock catalogues (as explained in the last section via Fig. \ref{fig: mostmassive}) which result from different clustering mechanisms occurring in different models (see Section \ref{sect:5b}). 
In addition, the mean HGC density shown in Fig \ref{fig: obs-histo}, in all cases is close to $1\times 10^{16}M_\odot / \text{ Mpc}^3$. Likewise, in (b) and (c) we show the distributions for different HGC samples in the density-crossing ($\rho_g-t_c$) time and density-velocity dispersion ($\rho_g-\sigma_v$) planes, respectively.
 
It is well known in the literature that the crossing time of a group, corresponds to a rough measure of the dynamical stability of such a system and can be defined as follows \citep{hickson-1992} 

\begin{equation}
 t_{c}=\frac{4}{\pi}\frac{R}{\sigma_v},
\end{equation}

where $R$ is the median projected separation between members and $\sigma_v$ is the mean total velocity dispersion of the group. In (b) we observe that the distributions for HCG in simulations partially match those of observed HCGs. Values of the cross-time below $3\times 10^{-1}H_0^{-1}$ are not consistent with those in all simulations, but only values above that threshold. In other words, these observed groups with small $t_{c}$ cannot be reproduced in any simulation considered here. 
A similar situation is evident in (c) where the velocity dispersion distribution for observed HCGs is much higher than those for HGC in simulations. More precisely, the mean values for the velocity dispersion in CDM, GDM I, GDM II and GDM III distributions are 69.9, 73.2, 68.6 and 89.7 $\text{km}/s$ respectively. On the other side, the same quantity for the observational HCGs is 344 $\text{km}/s$. 
 In conclusion, such a large discrepancy cannot be solved either within CDM or GDM and it is linked with one of the most interesting open problems regarding HCG \citet{Wiens_2019,Gimenez-Mamon}. 
 
It is important to mention that the best-fit for $\rho_{\text{g}}$ inferred for GDM III is the closest to the central observational value in comparison to the best-fit for other models A similar situation happens for the virial mass, where GDM scenarios are closer to the observation predictions. 

On the other side, the mean values for the crossing time for CDM, GDM I, GDM II and GDM III are $8.57\times 10^{-2}H_0^{-1}$, $7.43\times 10^{-2}H_0^{-1}$, $9.97\times 10^{-2}H_0^{-1}$ and $7.60\times 10^{-2}H_0^{-1}$ respectively and for the observations the value is $8.38\times 10^{-2}H_0^{-1}$. 
On the other hand, the parameter space velocity dispersion-crossing time is shown in \ref{fig: obs-histo} (d). According to the previous comments we can see the discrepancies between the velocity dispersion.
Since $t_c$ closely depends on the dynamics in HCGs, it is linked to the kinematics in such systems. Therefore, the existing tension for the velocity dispersion between observations and simulations is intimately related to that for the crossing time.

The HCG distribution in the mass-crossing time and mass-velocity dispersion planes shown in panels (e) and (f) of Figs. \ref{fig: obs-histo} respectively. 
For the mass, there are regions where the distributions for observations and simulations match. However, in the last panel (f), distributions for observations and models are disjoint but we can identify some points lying inside the simulated regions.

Another interesting fact is the correlations between different observables can be noticed in every panel. Some of them are more prominent than others, for example, panel (d) clearly shows that there exists a linear scaling relation between the crossing time and the velocity dispersion either in models or observations. Evidently, the slope corresponding to observations differs from that for simulations. This fact might bring some hints about the differences between the dynamics of real and simulated groups. In contrast in panel (f) a similar correlation can be noticed between the virial mass and the velocity dispersion, however, in this case, the slopes corresponding to models and observations are pretty similar. Despite the discrepancy between estimates of the velocity dispersion in observations and simulations, it is interesting that the $M_{\text{vir}}-\sigma_v$ scaling relation is the similar in both cases.

\subsection{Time Evolution of simulated HCGs}
\label{sec:6a}
The evolution of the physical and geometric properties of HCGs populations through time can be the key to understanding their dynamics. There are different works reported in the literature showing the impact of the environment on the evolution of the embedded galaxies \citep{Mendes-Oliveiera-1998,Mendes-Oliveiera-2005,Coenda,tovmassian-2006}. 
Nevertheless, catalogues of observed HCGs comprehend only regions in the nearby universe, redshift values reported in catalogues for observational HCGs are within a range ($7\times 10 ^{-3}$,$1.4\times 10^{-1}$). Consequently, a host of works studying CGs at high redshift by means of simulations have been developed in the last decade.

\begin{figure}
 \centering
 \begin{tabular}{c}
 \includegraphics[width=8cm]{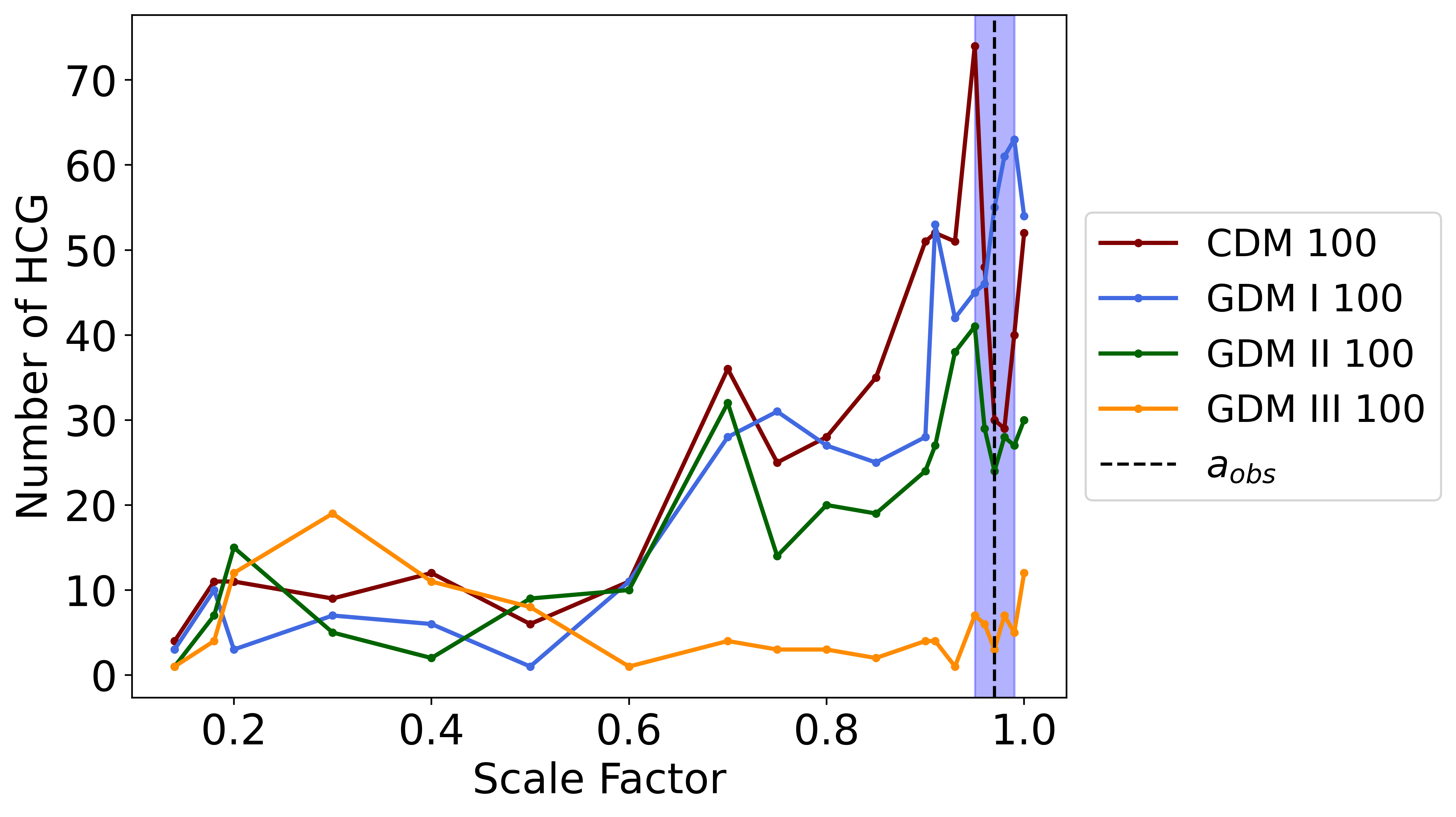}\\
 \textbf{(a)}\\
 \includegraphics[width=8cm]{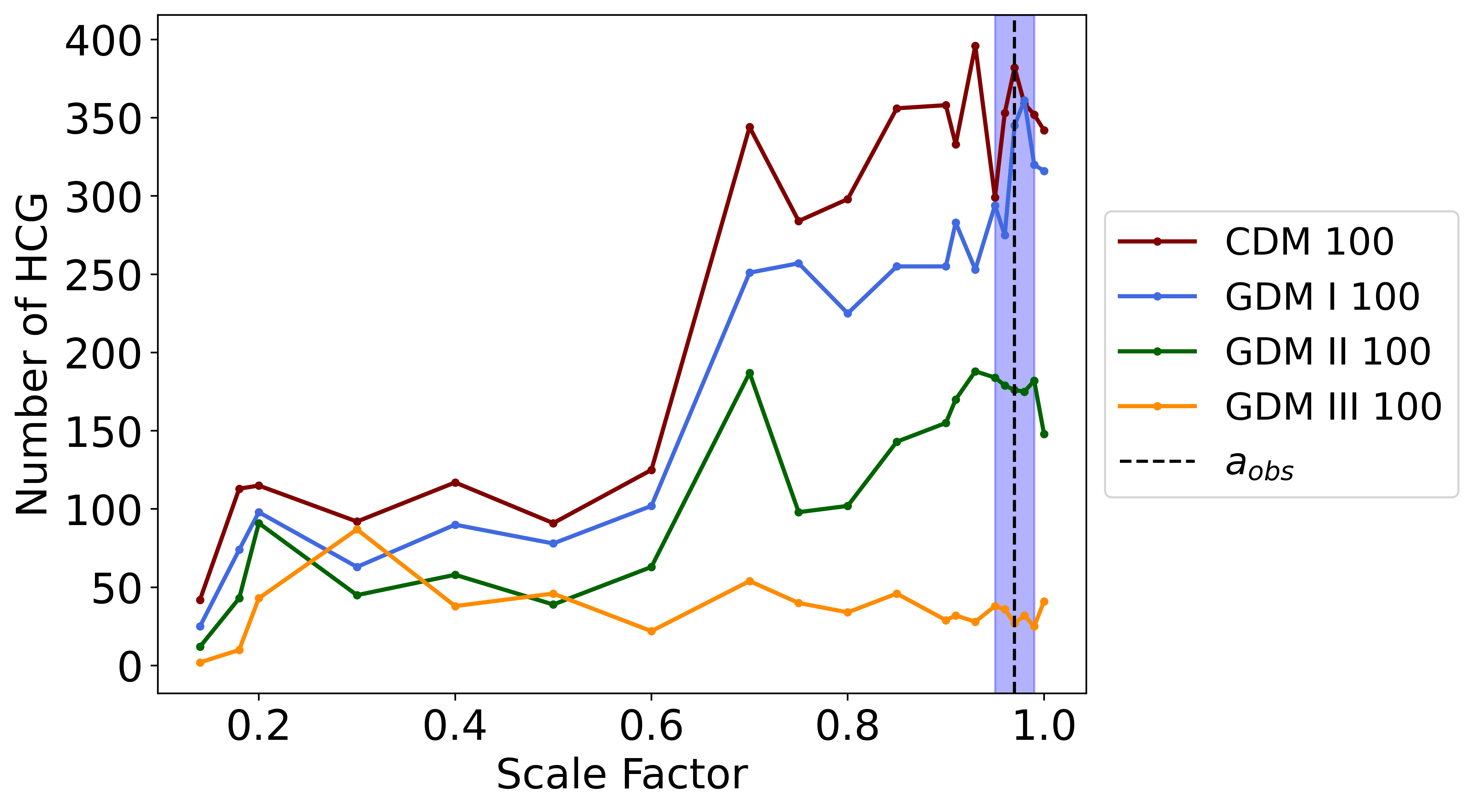}\\
 \textbf{(b)}
 \end{tabular}
 \caption{Time series of the number of HCGs through time for $N=4$ (a) and $N=3$ (b) in different GDM scenarios. The number of groups remains almost constant and very similar between models for a while ($a\in(0.2, 0.6)$). After this point, the HCG number increases significantly for CDM, GDM I and GDM II. The maximum value is reached at $a\sim 0.97$ in the three cases. On the other hand, GDM III seems to reach its maximum at $a\sim 0.3$. The bluish band and the central dashed line represent the median redshift of the observational sample and its uncertainty respectively. }
 \label{fig:sfvsN}
\end{figure}

The first goal in this work in this direction is to obtain the HCGs counts along the time for different models. With that purpose, we applied directly the algorithm of classification to our mock catalogues at different values of the scale factor corresponding to scale factors from a=0.14 to a=1.
The results are shown in Fig, \ref{fig:sfvsN} for N=4 (a) and N=3 (b). This plot shows that a maximum number of HCG is achieved at $a_{max}$ right before a=1. Our estimate for $a_{max}$ is close to the mean value corresponding to the observational HCGs sample represented by a dashed line. In both cases, the curve for the number of HCGs remains almost flat between a=0.2 and a=0.6, except for GDM III, which has the maximum value around a=0.3. After that, the number of groups starts to grow more efficiently within all the models, except for GDM III, whose counts remain around the same small value over time.
Another interesting feature of the time series of the counts is the existence of a local maximum at $a=0.7$ which arises just after the onset of the growing epoch at a=0.6. For some models (such as GDM II) the number of HCG at that point is comparable to the absolute maximum at $a_{max}$. This brings new questions motivating future works regarding HCGs evolution indeed.

The study of the evolution of each galaxy member in the CGs through time becomes more important. As a perspective, we aim to trace back in time the properties of galaxy members of groups up to z=0, also to characterise the environment in which they are embedded and as well as the interactions with neighbour galaxies. Besides, we would like to reconstruct the trajectory in the space-time of every member to study the mechanisms of HCGs clustering in different GDM models and environments.

\section{Discussion and conclusions}
\label{sec:7}

The main result of this work, is the study of some properties of populations of HCG in catalogues generated from N-body simulations with different initial seeds prescribed according to the linear matter power spectrum of three GDM models. Within this theoretical framework, the dark matter corresponds to a general fluid with non-vanishing sound speed, viscosity and equation of state. The GDM models chosen in this work have fixed and sufficiently small values of those free functions Consequently, dark matter linear perturbations free-stream giving rise to a cut-off in the matter power spectrum at small scales. Basically, an important goal of this work is to study the structure formation within these scenarios considering initial conditions with a lack of small substructure and its impact on counts and physical properties of populations of HCGs. 

As a first step, we needed to create mock galaxy catalogues for different models. We found important differences in the Halo and Stellar Mass Functions at small and intermediate scales which are more pronounced at high redshift. These differences are clearly linked to the lack of substructure from which the hierarchical formation of structure goes on. As a matter of fact, the large structures are created at redshifts nearby zero within all the scenarios considered.

A second important result was the implementation of an algorithm to identify HCGs in our mock catalogues by accommodating the original Hickson compactness criterion ( which uses projected separations between members) in terms of actual 3D physical separations available in mock catalogues. Specifically, we establish the compactness criterion by determining the ratio between physical and projected distances between members, namely 1.28. A very similar value was reported in \citet{Wiens_2019} estimated using a different method. Afterwards, by monitoring the impact of different filters onto counts of clusters we identified the most relevant corresponding to the isolation condition and the mass-ratio relation. The former mainly affects counts within models with a larger amount of substructure which prevents isolation of structures.
Besides, the mass-ratio condition mainly affects counts of models with less amount of substructure since in that case galaxy clusters are less likely to form given that massive galaxies are numerous. 
The main relevance of this criterion is that there are different galaxy clustering mechanisms in different models, in this work we identified self agglomeration (mainly for CDM) and clustering induced by a falling into a dominant gravitational
potential well (as in GDM III). 

Additionally, we computed the number of HCGs in populations corresponding to different GDM and CDM scenarios. We set the minimum number of neighbours within a cluster to $N=4$. As a complement, we consider $N=3$ in order to figure out how the counts of HCG change when this selection criterion is modified.

An interesting result is that, when the standard algorithm of classification with $N=4$ is used, a larger amount of HCG are counted in GDM I simulation in comparison to CDM counts even though the first one has less structure. Also, GDM II has a very close number of groups to CDM. However, GDM III shows a very small number since the beginning of the classification. This shows that HCGs can proliferate within GDM models despite the suppressed amount of small structures, which suggest modification either in the features of populations of field galaxies and either in the way on how these galaxies cluster to form HCG within models where the dark matter component is not perfectly cold or may have some extent of viscosity within the fluid limit.

For $N=3$ there is a different behaviour. First of all, the number of groups increases significantly. In addition, CDM is the dominant model closely followed by GDM I and now GDM II and III are well below the previous models.

In addition, we made a comparison of the properties of simulated groups and the observational ones in all models. The 1D and 2D histograms show a good agreement for the mass and the density in all the cases. The main differences arise for the velocity dispersion, where that for observed HCGs shows a much higher value than all the simulated models. These discrepancies have been reported in other works such as \citet{Wiens_2019, Gimenez-Mamon} and there is still no explanation about this situation. Nevertheless, the crossing-time graphics have overlapping regions between simulated and observed data.

Finally, in order to track the time evolution of counts of HCG in different populations, we applied the algorithm of classification to clustering in the mock catalogues from $z=0$ to $z=6$. The time series of the counts present an absolute maximum for CDM, GDM I and GDM II, whose value is quite similar to that reported for observed HCG and occurs at around the same redshift. A remarkable feature of the time series for such models is that it remains flat for a while before starting to significantly grow until reaching the absolute maximum close to $z=0$. In contrast, counts for HCGs in GDM III remain small and similar to the initial value. In general, this model shows the most notable differences in all tests here reported. In the future, we aim to carry out a more complete study of the evolution of groups and it would include the evolution of each member over time, as well as the study of their environment. These can be useful to understand the dynamic properties of the groups at the present and future time.

\section{Acknowledgements}
\label{sec:8}

For this research work, we use the NASA/IPAC Extragalactic Database (NED), which is operated by the Jet Propulsion Laboratory, California Institute of Technology, under contract with the National Aeronautics and Space Administration.

The authors thankfully acknowledge computer resources, technical advice and support provided by Laboratorio Nacional de Superc{\'{o}}mputo del Sureste de M{\'{e}}xico (LNS), a member of the CONACYT national laboratories, with project No. 202004054c.

The authors acknowledge VIEP and CIFFU for the financial support through an internal research project.
 J.L.S and E.M.V acknowledge CONACYT for the PhD scholarships and A.A.L. and O.M.B. Acknowledge CONACYT for the SNI financial support.

\bibliographystyle{mnras}
\bibliography{bibliografia.bib}


\appendix
\section{Tables}
\begin{table}
    \centering
    \begin{tabular}{l|c|c|c}
     & \textbf{$\text{log}_{10}$ Median} & \textbf{$\text{log}_{10}$ Mean} & \textbf{$\text{log}_{10}$ Mode}\\
    
     CDM & 12.1 & 12.4 & 11.6 \\
     GDM I & 12.4 & 12.6 & 12.0 \\
     GDM II & 12.6 & 12.7 & 12.2 \\
     GDM III & 12.6 & 13.5 & 11.9 \\
     Observations & 12.3 & 12.4 & 11.6
    \end{tabular}
    \caption{Logarithm of the median, mean and mode values for the virial mass in units of $M_{\text{sun}}$ of the HCG members in the simulations.}
    \label{tab:stat_mass}

\end{table}

\begin{table}
    \centering
    \begin{tabular}{l|c|c|c}
     & \textbf{$\text{log}_{10}$ Median} & \textbf{$\text{log}_{10}$ Mean} & \textbf{$\text{log}_{10}$ Mode}\\
    
     CDM & -1.2 & -1.0 & -1.0 \\
     GDM I & -1.2 & -1.1 & -1.2 \\
     GDM II & -1.2 & -1.0 & -1.1 \\
     GDM III & -1.1 & -1.1 & -1.1 \\
     Observations & -1.5 & -1.1 & -1.3
    \end{tabular}
    \caption{Logarithm of the median, mean and mode values for the crossing time in units of $H_0^{-1}$ of the HCG members in the simulations.}
    \label{tab:stat_tcr}
\end{table}

\begin{table}
    \centering
    \begin{tabular}{l|c|c|c}
     & \textbf{$\text{log}_{10}$ Median} & \textbf{$\text{log}_{10}$ Mean} & \textbf{$\text{log}_{10}$ Mode}\\
    
     CDM & 1.8 & 1.8 & 1.6 \\
     GDM I & 1.8 & 1.9 & 1.7 \\
     GDM II & 1.8 & 1.8 & 1.6 \\
     GDM III & 1.9 & 2.0 & 1.7 \\
     Observations & 2.6 & 2.6 & 2.2
    \end{tabular}
    \caption{Logarithm of the median, mean and mode values for the velocity dispersion in units of km/s of the HCG members in the simulations.}
    \label{tab:stat_veldisp}
\end{table}

\begin{table}
    \centering
    \begin{tabular}{l|c|c|c}
     & \textbf{$\text{log}_{10}$ Median} & \textbf{$\text{log}_{10}$ Mean} & \textbf{$\text{log}_{10}$ Mode}\\
    
     CDM & 15.3 & 16-0 & 14.3 \\
     GDM I & 15.6 & 16.1 & 14.8 \\
     GDM II & 15.5 & 16.1 & 15.0 \\
     GDM III & 15.9 & 16.5 & 14.8 \\
     Observations & 16.1 & 16.5 & 15.2
    \end{tabular}
    \caption{Logarithm of the median, mean and mode values for the mass density in units of $M_{\text{sun}}/\text{Mpc}^3$ of the HCG members in the simulations.}
    \label{tab:stat_density}
\end{table}

\bsp	
\label{lastpage}
\end{document}